\newif\ifshortver
\shortverfalse

\newif\iflongver

\ifshortver

\documentclass[onecolumn]{IEEEtran}
\longverfalse

\newcommand{\myskip}{}
\newcommand{\mysmallskip}{}

\else

\documentclass[english,onecolumn]{IEEEtran}
\longvertrue

\newcommand{\myskip}{\medskip}
\newcommand{\mysmallskip}{\medskip}

\fi

\IEEEoverridecommandlockouts

\usepackage[utf8]{inputenc} 
\usepackage[T1]{fontenc}
\usepackage{url}
\usepackage{ifthen}
\usepackage{cite}
\usepackage[cmex10]{amsmath} 

\usepackage{amsmath}
\usepackage{amssymb}
\usepackage{graphicx}
\usepackage{esint}

\makeatletter
\def\useAmsThmOrIEEE{0}
\ifnum\useAmsThmOrIEEE=1
\usepackage{amsthm}

\theoremstyle{plain}
\newtheorem{thm}{\protect\theoremname}
\theoremstyle{definition}
\newtheorem{defn}[thm]{\protect\definitionname}
\theoremstyle{plain}
\newtheorem{prop}[thm]{\protect\propositionname}
\theoremstyle{plain}
\newtheorem{lem}[thm]{\protect\lemmaname}
\theoremstyle{plain}
\newtheorem{cor}[thm]{\protect\corollaryname}
\theoremstyle{definition}
\newtheorem{example}[thm]{\protect\examplename}
\theoremstyle{definition}
\newtheorem{rem}[thm]{\protect\remarkname}
\else

\newtheorem{thm}{\protect\theoremname}
\newtheorem{defn}[thm]{\protect\definitionname}
\newtheorem{prop}[thm]{\protect\propositionname}

\newtheorem{cor}[thm]{\protect\corollaryname}

\newtheorem{rem}[thm]{\protect\remarkname}
\fi

\DeclareMathOperator{\sinc}{sinc}

\usepackage{mathrsfs}
\usepackage{algorithm,algpseudocode}

\usepackage{booktabs}    
\usepackage{siunitx}       
\usepackage{multirow}      
\usepackage{colortbl}
\definecolor{lightgray}{rgb}{0.9,0.9,0.9}
\definecolor{lightred}{rgb}{1,0.8,0.8}
\definecolor{lightgreen}{rgb}{0.6,1,0.6}
\definecolor{lightyellow}{rgb}{1,1,0.5}
\definecolor{lightgrey}{rgb}{0.8,0.8,0.8}

\IEEEoverridecommandlockouts

\allowdisplaybreaks[1]

\makeatother

\usepackage{babel}
\providecommand{\corollaryname}{Corollary}
\providecommand{\definitionname}{Definition}
\providecommand{\propositionname}{Proposition}
\providecommand{\theoremname}{Theorem}
\providecommand{\lemmaname}{Lemma}
\providecommand{\remarkname}{Remark}

\begin{document}
\title{Rejection-Sampled Universal Quantization for Smaller Quantization Errors}
\author{
Chih Wei Ling and Cheuk Ting Li
\thanks{
This work was partially supported by two grants from the Research
Grants Council of the Hong Kong Special Administrative Region, China
{[}Project No.s: CUHK 24205621 (ECS), CUHK 14209823 (GRF){]}.

This paper was presented in part at the 2024 IEEE International Symposium on Information Theory (ISIT).

Chih Wei Ling was with the Department of Information Engineering, The Chinese University of Hong Kong, Hong Kong SAR of China. He is now with the Department of Computer Science, The City University of Hong Kong, Hong Kong SAR of China.
Cheuk Ting Li is with the Department of Information Engineering, The Chinese University of Hong Kong, Hong Kong SAR of China. Email: ctli@ie.cuhk.edu.hk.

}
}

\maketitle
\begin{abstract}
We construct a randomized vector quantizer which has a smaller maximum error compared to all known lattice quantizers with the same entropy for dimensions 5, 6, ..., 48, and also has a smaller mean squared error compared to known lattice quantizers with the same entropy for dimensions 35, ..., 47, in the high resolution limit. Moreover, our randomized quantizer has a desirable property that the quantization error is always uniform over the ball and independent of the input. Our construction is based on applying rejection sampling on universal quantization, which allows us to shape the error distribution to be any continuous distribution, not only uniform distributions over basic cells of a lattice as in conventional dithered quantization. We also characterize the high SNR limit of one-shot channel simulation for any additive noise channel under a mild assumption (e.g., the AWGN channel), up to an additive constant of 1.45 bits. 
\end{abstract}

\myskip

\section{Introduction}
Entropy-constrained quantization has been extensively studied in the past decades.
For scalar quantization,
it was shown by Gish and Pierce \cite{gish1968asymptotically} that the uniform quantizer, followed by entropy coding, is asymptotically optimal under certain assumptions.
They also compared the entropy of the uniform quantizer with the rate-distortion function \cite{shannon1959coding} for some difference distortion measures.
In \cite{zador1982asymptotic} (also see \cite{Gray2002onZador}), Zador studied the trade-off between the distortion and the rate for vector quantizers, for both fixed-rate quantizers and entropy-constrained quantizers.
Gersho \cite{gersho1979asymptotically} generalized the study in \cite{gish1968asymptotically} to vector quantizers and conjectured that the tessellating quantizers, which include lattice quantizers, are asymptotically optimal.
Chou et al. \cite{chou1989entropy} studied entropy-constrained vector quantizers by formulating it as a Lagrangian model similar to Lloyd's \cite{lloyd1982least} and designed a greedy algorithm to find the local optimal vector quantizer.

A randomized quantizer is a quantizer that depends on a random state independent of the input signal. A popular example is \emph{subtractive dithering} 
\iflongver
\cite{roberts1962picture,Limb1969visual,Jayant1972speech}.
\else
\cite{roberts1962picture}.
\fi
Entropy-constrained quantization is often used in conjunction with randomized quantizers, such as dithered quantization. 
Notably, it was shown in the works on \emph{universal quantization} by Ziv \cite{ziv1985universal} and Zamir and Feder \cite{zamir1992universal} that 
dithered quantization has a (conditional) entropy within a constant redundancy away from the rate-distortion function for difference distortion measures. See~\cite{linder1994asymptotic,zamir1996lattice} for further analyses on universal quantization.
Dithered quantization was combined with channel simulation techniques to create a nonuniform error distribution in~\cite{theis2022algorithms}.

The maximum error (or maximum distortion) of a quantizer (i.e., the largest possible magnitude of the error)
has also been studied. For lattice quantizers, the maximum error is given by the covering radius~\cite{rogers1964packing}. The maximum error has also been studied in epsilon entropy~\cite{Kolmogorov1956epsilon,posner1971epsilon} and $D$-semifaithful codes~\cite{ornstein1990universal,zhang1997redundancy,kontoyiannis2000pointwise}.
Notably, $D$-semifaithful codes are usually entropy-constrained (variable-length) instead of fixed-length to ensure that the distortion is bounded almost surely.

In this paper, which is the complete version of \cite{Ling2024ISITRSUQ},\footnote{The conference paper \cite{Ling2024ISITRSUQ} includes the descriptions (but not the complete proof) of the rejection-sampled universal quantizer (RSUQ) for error uniformly distributed over an arbitrary set, and the layered rejection-sampled universal quantizer (LRSUQ) for general (uniform or nonuniform) continuous error distribution. Compared to the conference version \cite{Ling2024ISITRSUQ}, this complete version includes the complete analysis on the RSUQ and the LRSUQ (such as Theorems \ref{thm:layered_alt}, \ref{thm:rsuq_ent_layer} and \ref{thm:nonunif_converse} that are not included in \cite{Ling2024ISITRSUQ}), 
the Zador redundancy~\eqref{eq:red_mse_zador},
a new result in Theorem~\ref{thm:rsuq_ent_gen} about the upper and lower bounds on the conditional entropy of the RSUQ in terms of its normalized entropy and the differential entropy of a given source when the source follows a nonuniform distribution that is ``sufficiently spread out'', a corollary to Theorem~\ref{thm:rsuq_ent_gen} about the characterization of the conditional entropy of the RSUQ in the high-resolution limit, and discussions on the normalized excess information of LRSUQ.} we construct a randomized vector quantizer, called \emph{rejection-sampled universal quantizer} (RSUQ), based on applying rejection sampling on top of a universal quantizer, in order to guarantee that the error follows an arbitrary uniform distribution or an arbitrary continuous distribution (using the layered construction in~\cite{wilson2000layered,hegazy2022randomized}). Main contributions:
\begin{itemize}
\item The RSUQ, with an error distribution uniform over the $n$-ball, has a smaller maximum error compared to all known lattice quantizers with the same entropy for dimensions $n=5,6,\ldots,48$, showing that RSUQ performs better than known lattice quantizers as a $D$-semifaithful code, and it also has a smaller mean squared error compared to known lattice quantizers with the same entropy for dimensions $n=35,\ldots,47$, in the high resolution limit.
\item The RSUQ has a redundancy at most $(\log_2 e)/n$ bits per dimension from the rate-distortion lower bound for a given maximum error in the high resolution limit, smaller than the $O((\log_2 n)/n)$ bound on the redundancy for lattice quantizers with respect to maximum error~\cite{rogers1964packing}. 
\item The RSUQ also has a redundancy at most $(\log_2 e)/n$ bits per dimension from Zador's lower bound \cite{zador1982asymptotic} for a given mean squared error in the high resolution limit, smaller than the recent bound $(n^{-1}+4n^{-2}+8n^{-3})\log e$ for lattice quantizers by Ordentlich \cite{ordentlich2025voronoi}. 
\item Unlike conventional lattice quantizers which are often difficult to construct for large $n$, RSUQ can be constructed over any lattice (e.g., $\mathbb{Z}^n$), and its error is not sensitive to the choice of lattice. This greatly simplifies the construction of RSUQ.
\item  
For nonuniform source distribution, we bound the conditional entropy of the RSUQ in terms of the differential entropy of the source, under certain smoothness conditions. This is similar to the result obtained by Linder and Zeger  \cite{linder1994asymptotic} for conventional lattice quantizers.
\item We also prove that RSUQ has a similar universal quantization property as \cite{ziv1985universal,zamir1992universal}, in the sense that it has an entropy within a constant gap away from the rate-distortion function regardless of the source distribution.
\item We also study the entropy required to ensure that the quantization error is precisely a prescribed continuous distribution (i.e., we are simulating an additive noise channel~\cite{bennett2002entanglement,sfrl_trans,hegazy2022randomized}), where we characterize the optimal entropy in the ``high SNR limit'' (the range of the input signal tends to infinity) within a constant gap of $\log_2 e$.
\item We also discuss the normalized excess information of a randomized quantizer that exactly simulates a given channel. 
This quantity is defined as the difference, in the high-SNR limit, between the conditional entropy of the randomized quantizer and the mutual information between the input and output signals of the channel, by analogy to the redundancy \cite{zamir1992universal} in the lossy compression setting. 
Additionally, we also compare the normalized excess information of LRSUQ with other randomized quantizers, such as the layered shift-periodic quantizer in \cite{ling2023vector} and the rotated dithered quantizer in \cite{KobusRotated2024}.
\end{itemize}

\ifshortver
Proofs are in the preprint \cite{ling2024rejectionsampled} due to space constraint.
\fi

\subsection{Related Works}

Randomized compression where the reconstruction follows a prescribed conditional distribution given the source is referred to as \emph{channel simulation}~\cite{bennett2002entanglement,winter2002compression,cuff2013distributed}. RSUQ in this paper can be regarded as a one-shot channel simulation scheme for simulating an additive noise channel with an arbitrary continuous noise distribution. Compared to existing one-shot channel simulation schemes~\cite{harsha2010communication,sfrl_trans} with a logarithmic optimality gap, RSUQ only has a constant optimality gap 
\ifshortver
(see \cite{ling2024rejectionsampled}).
\else
(Corollary~\ref{cor:rsuq_exist}). 
\fi
Additive noise channel simulation has applications to machine learning \cite{havasi2019minimal,agustsson2020universally,flamich2020compressing} and differential privacy 
\iflongver
\cite{lang2022joint,shahmiri2023communication,hasirciouglu2024communication,hegazy2023compression,yan2023layered,hegazy2024compression}, 
\else
\cite{lang2022joint,shahmiri2023communication,hasircioglu2023communication,hegazy2023compression}, 
\fi
and RSUQ may be applied to these areas. 
For a comprehensive treatment of the subject, we direct the interested reader to the monograph in \cite{Li2024ChannelSim}.

The greedy rejection sampling scheme by Harsha \textit{et al.} \cite{harsha2010communication} is a channel simulation scheme based on rejection sampling, where the acceptance probability can change at each iteration.
In \cite{flamich2023adaptive}, Flamich and Theis generalized greedy rejection sampling \cite{harsha2010communication} to also allow the proposal distribution to change at each iteration. 
Their scheme can be combined with a variant of scalar dithered quantization based on bits-back coding \cite{bamler2022understanding} to perform channel simulation for one-dimensional Gaussian noise channel. In contrast, the layered rejection-sampled universal quantizer in this paper is based on \emph{vector} dithered quantization, and works for general error distributions of any dimension.

The original and the greedy rejection sampling schemes are examples of causal rejection sampling algorithms (CRS) \cite{liu2018rejection, goc2024channel}. In \cite{goc2024channel}, Goc and Flamich analyzed the performance of CRS algorithms in terms of the \emph{channel simulation divergence}, a new quantity related to the \emph{layered entropy} defined by Hegazy and Li \cite{hegazy2022randomized}.
The discrete counterpart of the layered entropy has been defined and studied in depth in \cite{Li2025discreteLE}.
Sriramu and Wagner \cite{Sriramu2024OptimalRI} recently proposed a two-stage rejection sampling scheme for asymptotic exact channel simulation, which achieves the optimal asymptotic redundancy.
Very recently, Flamich, Sriramu, and Wagner \cite{flamich2025NonSingularChan} have completed the characterization of redundancy in channel simulation algorithms.

A related work \cite{theis2022algorithms} combined dithered quantization with \emph{ordered random coding} (which combines 
\iflongver
two channel simulation techniques: 
\fi
\emph{minimal random coding}~\cite{havasi2019minimal} and \emph{Poisson functional representation}~\cite{sfrl_trans,li2021unified}) to create a nonuniform error distribution. Our construction combines dithered quantization with \emph{rejection sampling} and \emph{layered quantization}~\cite{wilson2000layered,hegazy2022randomized} instead. Unlike \cite{theis2022algorithms} which focuses on channel simulation, our proposed scheme also works well as a compression scheme, ensuring lower error than known lattice quantizers for some situations.
We note that the RSUQ scheme in this paper coincides with the hybrid coding scheme in \cite{theis2022algorithms} for the special case where the lattice used is the integer lattice, and the noise distribution is a uniform distribution. For the simulation of additive noise channels with nonuniform noise distribution, the layered RSUQ scheme in this paper is rather different from the channel simulation scheme in \cite{theis2022algorithms}.\footnote{\cite[Theorem 3.5]{theis2022algorithms} requires the support of the noise distribution to be inside a hyperrectangle, and the bound on the communication cost depends on the size of the hyperrectangle. In comparison, LRSUQ in this paper can be applied to noise distributions with unbounded support, and has an almost-optimal communication cost in the high-SNR limit.}

A recent work \cite{ling2023vector} also studied a construction, called \emph{shift-periodic quantizers}, which can ensure that the error is uniform over any prescribed set. The main difference between \cite{ling2023vector} and our construction is that, to ensure a uniform error over $\mathcal{A}\subseteq \mathbb{R}^n$, \cite{ling2023vector} decomposes a basic cell of a lattice of the same volume as $\mathcal{A}$ into infinitely many pieces to form $\mathcal{A}$, whereas the RSUQ in this paper starts with a basic cell containing $\mathcal{A}$ and performs rejection sampling. 
Compared to \cite{ling2023vector}, RSUQ is significantly simpler to implement,\footnote{The construction in \cite{ling2023vector}, 
even for $n=2$, is already quite hard to characterize.} and requires a smaller entropy for every dimension $n\ge 2$ (see Figure \ref{fig:compare_lattice}).

Recently, Kobus et al. \cite{KobusRotated2024} introduced a computationally-efficient scheme involving a rotated dithered lattice quantizer to approximately simulate a Gaussian channel.
Nevertheless, the quantization error in their scheme only approximately follows a Gaussian distribution, whereas the LRSUQ scheme in our paper (Section~\ref{sec:LRSUQ}) ensures a precise noise distribution.

\mysmallskip

\subsection*{Notations}

Entropy is in bits, and logarithms are to the base $2$. 
Write $H(X)$ for the entropy, and $h(X)$ for the differential entropy. For $\mathcal{A},\mathcal{B}\subseteq\mathbb{R}^{n}$,
$\gamma\in\mathbb{R}$, $\mathbf{G}\in\mathbb{R}^{n\times n}$, $\mathbf{z}\in\mathbb{R}^{n}$
write $\gamma\mathcal{A}:=\{\gamma\mathbf{x}:\,\mathbf{x}\in\mathcal{A}\}$,
$\mathbf{G}\mathcal{A}:=\{\mathbf{G}\mathbf{x}:\,\mathbf{x}\in\mathcal{A}\}$,
$\mathcal{A}+\mathbf{z}:=\{\mathbf{x}+\mathbf{z}:\,\mathbf{x}\in\mathcal{A}\}$,
$\mathcal{A}+\mathcal{B}=\{\mathbf{x}+\mathbf{y}:\,\mathbf{x}\in\mathcal{A},\,\mathbf{y}\in\mathcal{B}\}$
for the Minkowski sum, $\mathcal{A}-\mathcal{B}=\{\mathbf{x}-\mathbf{y}:\,\mathbf{x}\in\mathcal{A},\,\mathbf{y}\in\mathcal{B}\}$,
and $\mu(\mathcal{A})$ for the Lebesgue measure of $\mathcal{A}$.
Let $B^{n}:=\{\mathbf{x}\in\mathbb{R}^{n}:\,\Vert\mathbf{x}\Vert\le1\}$
be the unit $n$-ball, and let $\kappa_{n}:= \mu(B^n) = \pi^{n/2}/\Gamma(n/2+1)$
be the volume of the $n$-ball.

\myskip

\section{Randomized Quantizers}

The definition of a randomized quantizer, which includes dithered
quantization \cite{roberts1962picture,ziv1985universal,zamir1992universal}
and distribution preserving quantization \cite{li2010distribution}
as special cases, is given below.

\myskip

\begin{defn}
\label{def:rand_quant}A \emph{randomized quantizer} is a pair $(P_{S},Q)$,
where $P_{S}$ is a distribution over $\mathcal{S}$ (the \emph{state
distribution}), $Q:\mathbb{R}^{n}\times\mathcal{S}\to\mathbb{R}^{n}$
is the \emph{quantization function} satisfying that the set $Q(\mathbb{R}^{n},s)=\{Q(\mathbf{x},s):\,\mathbf{x}\in\mathbb{R}^{n}\}$
is finite or countable for every $s\in\mathcal{S}$. In particular,
if $P_{S}$ is a degenerate distribution, we call this a \emph{deterministic
quantizer}.
\end{defn}
\medskip{}

Operationally, we first generate the random state $S\sim P_{S}$ which
is available to both the encoder and the decoder. The encoder observes
$S$ and a random input $\mathbf{X}\in\mathbb{R}^{n}$, computes $\mathbf{Y}=Q(\mathbf{X},S)$,
and compresses $\mathbf{Y}$ in a variable-length sequence of bits
and sends it to the decoder, which is possible since $\mathbf{Y}\in Q(\mathbb{R}^{n},S)$
has at most countably many choices given $S$. For example, the encoder
can compress $\mathbf{Y}$ using a Huffman code \cite{huffman1952method}
conditional on $S$. The decoder can then use $S$ to decode the sequence
of bits and recover $\mathbf{Y}$. One can also view $(Q(\cdot,s))_{s\in\mathcal{S}}$
as an ensemble of quantizer (each $Q(\cdot,s):\mathbb{R}^{n}\to\mathbb{R}^{n}$
is a quantizer), and one of the quantizers is selected at random (see
\cite{li2010distribution}).

The Huffman encoding of $Q(\mathbf{X},S)$ conditional on $S$ requires
an expected length approximately $H(Q(\mathbf{X},S)|S)$. Therefore,
as in the works on universal quantization \cite{ziv1985universal,zamir1992universal},
the conditional entropy $H(Q(\mathbf{X},S)|S)$ measures the average
number of bits needed. We are particularly interested in the growth
of $H(Q(\mathbf{X},S)|S)$ when $\mathbf{X}$ is uniform over a large
set. The following is a straightforward generalization of the quantity
studied in \cite{hegazy2022randomized} to $\mathbb{R}^{n}$, which
also coincides with the normalized entropy considered in \cite{ling2023vector}
for the special case of shift-periodic quantizers.

\myskip

\begin{defn}
\label{def:norment}The \emph{normalized entropy} of a randomized
quantizer $(P_{S},Q)$ 
\iflongver
is defined as 
\[
\overline{H}(Q):= \underset{\tau\to\infty}{\lim\sup}\big(H(Q(\mathbf{X}_{\tau},S)|S)-\log\mu(\tau B^n)\big),
\]
where $\mathbf{X}_{\tau}\sim\mathrm{Unif}(\tau B^n)$ is independent of
$S\sim P_{S}$.
\else
is (letting $\mathbf{X}_{\tau}\sim\mathrm{Unif}(\tau B^n)$, $S\sim P_{S}$):
\[
\overline{H}(Q):= \underset{\tau\to\infty}{\lim\sup}\big(H(Q(\mathbf{X}_{\tau},S)|S)-\log\mu(\tau B^n)\big).
\]
\fi
\end{defn}
\medskip{}

The normalized entropy can also be regarded as a high resolution limit~\cite{gersho1979asymptotically,zamir1992universal}. If we quantize $\mathbf{X}\sim\mathrm{Unif}(a B^n)$ by a scaled version of the quantizer, i.e., $\mathbf{Y}_{\tau} = \tau^{-1} Q(\tau \mathbf{X},S)$, then $\lim\sup_{\tau\to\infty}\big(H(\mathbf{Y}_{\tau}|S)-\log\mu(\tau a B^n)\big) = \overline{H}(Q)$.

Note that the normalized entropy may be negative. Loosely speaking,
when $\mathbf{X}\sim\mathrm{Unif}(\mathcal{X})$ is uniform over a
large set $\mathcal{X}$, we expect $H(Q(\mathbf{X},S)|S)\approx\overline{H}(Q)+\log\mu(\mathcal{X})$.
Generally, when $\mathbf{X}$ has a nonuniform distribution that is sufficiently spread out, we expect 
\begin{align}
    & H(Q(\mathbf{X},S)|S)\approx\overline{H}(Q)+h(\mathbf{X}), \label{eqn:condent_diffent}
\end{align}
where $h(\mathbf{X})$ is the differential entropy of $\mathbf{X}$. This relation holds for conventional lattice quantizers in the high resolution limit (see \eqref{eqn:lattice_diffent}), and will be proved rigorously in Theorem \ref{thm:rsuq_ent_gen} for the randomized quantizer constructed in this paper.
Therefore, even though $\overline{H}(Q)$ is defined on a uniformly distributed input $\mathbf{X}$, it can describe the entropy of the quantizer for more general distributions of $\mathbf{X}$.

\iflongver
The error of the quantizer is $Q(\mathbf{X},S)-\mathbf{X}$,
and the magnitude of the error is $\Vert Q(\mathbf{X},S)-\mathbf{X}\Vert$.
\else
The error of the quantizer is $Q(\mathbf{X},S)-\mathbf{X}$.
\fi
The \emph{maximum error} of a randomized quantizer is the largest possible (magnitude of) error it can make:
\iflongver
\begin{align}
 & \sup_{\mathbf{x}\in\mathbb{R}^{n},\, s \in \mathcal{S}} \big\Vert Q(\mathbf{x},s)-\mathbf{x}\big\Vert. \label{eq:max_error}
\end{align}
\else
\begin{align}
 & {\sup}_{\mathbf{x}\in\mathbb{R}^{n},\, s \in \mathcal{S}} \, \Vert Q(\mathbf{x},s)-\mathbf{x}\Vert. \label{eq:max_error}
\end{align}
\fi

The following lower bounds of the normalized entropy are analogues of Shannon's rate-distortion lower bound~\cite{shannon1959coding} and Zador's lower bound~\cite{zador1982asymptotic} for randomized quantizers.
The proofs are included
in 
\ifshortver
the preprint
\cite{ling2024rejectionsampled}
\else
Appendix \ref{subsec:pf_lb} 
\fi
for completeness.

\mysmallskip

\begin{prop}
[Rate-distortion lower bounds]\label{prop:lb} For any randomized quantizer $(P_{S},Q)$,
we have the following lower bounds on the normalized entropy:
\begin{itemize}
\item If the maximum error \eqref{eq:max_error} is at most $r$, then (recall that $\kappa_n := \mu(B^n)$.)
\[
\overline{H}(Q) \ge -n\log r-\log\kappa_{n}.
\]
\item If the mean squared error 
satisfies (let $\mathbf{X}_{\tau}\sim\mathrm{Unif}(\tau B^n)$)
\[
\lim {\sup}_{\tau \to \infty}\mathbb{E}[\Vert Q(\mathbf{X}_{\tau},S)-\mathbf{X}_{\tau} \Vert^2] \le D,
\] 
then we have the following bound (which we call \emph{Shannon's lower bound})
\[
\overline{H}(Q) \ge -\frac{n}{2}\log(2\pi eD / n),
\]
and also the following tighter bound (which we call \emph{Zador's lower bound})
\[
\overline{H}(Q)\ge-\frac{n}{2}\log\frac{(n+2)D}{n}-\log\kappa_{n}.
\]
\end{itemize}
\end{prop}

\myskip

Given the rate-distortion lower bounds, we can define the \emph{redundancy} of a randomized quantizer $Q$ in a similar manner as~\cite{zamir1992universal}. The \emph{redundancy with respect to maximum error} (per dimension) is 
\begin{align}
\frac{1}{n}\overline{H}(Q) +\log r+\frac{1}{n}\log\kappa_{n}, \label{eq:red_maxerror}
\end{align}
which must be nonnegative,
where $r$ is the maximum error \eqref{eq:max_error} of $Q$. 
For the redundancy w.r.t. MSE, since there are two lower bounds in Proposition \ref{prop:lb}, we define two notions of redundancy:
the \emph{Shannon redundancy w.r.t. MSE} (per dimension) 
\begin{align}
\frac{1}{n}\overline{H}(Q) +\frac{1}{2}\log \frac{2\pi eD}{n}, \label{eq:red_mse_shannon}
\end{align}
and the \emph{Zador redundancy w.r.t. MSE} (per dimension) 
\begin{align}
\frac{1}{n}\overline{H}(Q) + \frac{1}{2}\log\frac{(n+2)D}{n}+\frac{1}{n}\log\kappa_{n}, \label{eq:red_mse_zador}
\end{align}
where 
$D=\lim {\sup}_{\tau \to \infty}\mathbb{E}[\Vert Q(\mathbf{X}_{\tau},S)-\mathbf{X}_{\tau} \Vert^2]$.
These redundancies are nonnegative due to Proposition \ref{prop:lb}. Since Zador's lower bound is tighter than Shannon's lower bound, \eqref{eq:red_mse_zador} is smaller and is a better measure of the optimality gap of a randomized quantizer compared to \eqref{eq:red_mse_shannon}. Nevertheless, we will also discuss \eqref{eq:red_mse_shannon} since redundancy is conventionally measured against Shannon's lower bound in the study of quantization \cite[Chapter 5]{zamir2014}. The gap between \eqref{eq:red_mse_shannon} and \eqref{eq:red_mse_zador} is $(\log n)/(2n)+O(1/n)$, implying that the Shannon redundancy w.r.t. MSE cannot be smaller than $(\log n)/(2n)+O(1/n)$; see Appendix~\ref{subsec:pf_mse_gap}.

\medskip{}

\section{Lattice Quantizers\label{sec:lattice}}

We review the basics of lattices, which can be found in \cite{conway2013sphere,zamir2014}.  Given a full-rank generator matrix
$\mathbf{G}\in\mathbb{R}^{n\times n}$ of a lattice $\mathbf{G}\mathbb{Z}^{n}=\{\mathbf{G}\mathbf{j}:\,\mathbf{j}\in\mathbb{Z}^{n}\}$,
its \emph{packing radius} is 
\ifshortver
\begin{align}
\underline{\lambda}(\mathbf{G}) & := (1/2)\min_{\mathbf{j}\in\mathbb{Z}^{n}\backslash\{\mathbf{0}\}}\Vert\mathbf{G}\mathbf{j}\Vert,\label{eq:pack_radius}
\end{align}
\else
\begin{align}
\underline{\lambda}(\mathbf{G}) & :=\sup\big\{\lambda\ge0:\,(\lambda B^{n}+\mathbf{G}\mathbf{j})_{\mathbf{j}\in\mathbb{Z}^{n}}\;\text{are disjoint}\big\}\nonumber \\
 & =\frac{1}{2}\min_{\mathbf{j}\in\mathbb{Z}^{n}\backslash\{\mathbf{0}\}}\Vert\mathbf{G}\mathbf{j}\Vert,\label{eq:pack_radius}
\end{align}
\fi
and its \emph{covering radius} is
\begin{align}
\overline{\lambda}(\mathbf{G}) & :=\inf\big\{\lambda\ge0:\,\lambda B^{n}+\mathbf{G}\mathbb{Z}^{n}=\mathbb{R}^{n}\big\}.\label{eq:cover_radius}
\end{align}
Its \emph{packing density} $\delta(\mathbf{G})$ and \emph{covering
density} $\Theta(\mathbf{G})$ are
\begin{equation}
\delta(\mathbf{G}):=\frac{(\underline{\lambda}(\mathbf{G}))^{n}\kappa_{n}}{|\det\mathbf{G}|},\;\Theta(\mathbf{G}):=\frac{(\overline{\lambda}(\mathbf{G}))^{n}\kappa_{n}}{|\det\mathbf{G}|}.\label{eq:density}
\end{equation}

A bounded set $\mathcal{P}\subseteq\mathbb{R}^{n}$ is called a \emph{basic
cell} of the lattice $\mathbf{G}\mathbb{Z}^{n}$ if $(\mathcal{P}+\mathbf{G}\mathbf{j})_{\mathbf{j}\in\mathbb{Z}^{n}}$
forms a partition of $\mathbb{R}^{n}$ \cite{conway2013sphere, zamir2014}. This implies
that $\mu(\mathcal{P})=|\det\mathbf{G}|$. In particular, the Voronoi
cell $\mathcal{V}:=\{\mathbf{x}\in\mathbb{R}^{n}:\,\arg\min_{\mathbf{j}\in\mathbb{Z}^{n}}\Vert\mathbf{x}-\mathbf{G}\mathbf{j}\Vert=\mathbf{0}\}$
(break ties in lexicographical order of $\mathbf{j}$) is a basic
cell. Given a basic cell $\mathcal{P}$, we can define a deterministic
quantizer $Q_{\mathcal{P}}:\mathbb{R}^{n}\to\mathbb{R}^{n}$, $Q_{\mathcal{P}}(\mathbf{x})=\mathbf{y}$
where $\mathbf{y}\in\mathbf{G}\mathbb{Z}^{n}$ is the unique lattice
point such that $\mathbf{x}\in-\mathcal{P}+\mathbf{y}$. The normalized
entropy (Definition \ref{def:norment}) of $Q_{\mathcal{P}}$ is $\overline{H}(Q) =-\log\mu(\mathcal{P})= -\log|\det\mathbf{G}|$.\footnote{This follows from Theorem \ref{thm:rsuq_ent} by substituting $\mathcal{A}=\mathcal{P}$.}
Moreover, under certain regularity conditions, given the basic cell $\mathcal{P}$ and some scaling factor $\alpha > 0$, the entropy of the lattice quantizer $Q_{\alpha \mathcal{P}}$ with the lattice partition $\{\alpha \mathcal{P} + \alpha\mathbf{G}\mathbf{v}:\, \mathbf{v} \in \mathbb{Z}^n\}$ when the input $\mathbf{X}$ follows a nonuniform distribution can be described as (see \cite[Equation (18)]{linder1994asymptotic} and \cite{Csiszar1973entropy})
\begin{align}
H(Q_{\alpha \mathcal{P}}(\mathbf{X})) = - \log \mu (\alpha \mathcal{P}) + h(\mathbf{X}) + o(1) \label{eqn:lattice_diffent}
\end{align}
as the scaling factor $\alpha \to 0$, i.e., in the ``high-resolution limit''. Hence, the relation \eqref{eqn:condent_diffent} holds.

Conventionally, we usually take $\mathcal{P}$ to be the Voronoi cell
$\mathcal{V}$. The resultant deterministic quantizer $Q_{\mathcal{V}}$
has a maximum error $\overline{\lambda}(\mathbf{G})$. The redundancy w.r.t. maximum error~\eqref{eq:red_maxerror} is\footnote{The redundancy w.r.t. maximum error $n^{-1} \log \Theta(\mathbf{G})$ for a lattice quantizer coincides with the logarithm of the \emph{covering efficiency}~\cite{zamir2014}.}
\ifshortver
\begin{align}
-\frac{1}{n}\log\mu(\mathcal{V}) +  \log \overline{\lambda}(\mathbf{G}) + \frac{1}{n}\log \kappa_n \; =\;  \frac{1}{n}\log \Theta(\mathbf{G}).\label{eq:lattice_norment}
\end{align}
\else
\begin{align}
& -\frac{1}{n}\log\mu(\mathcal{V}) +  \log \overline{\lambda}(\mathbf{G}) + \frac{1}{n}\log \kappa_n  \nonumber \\
& \;\; =  \frac{1}{n}\log \Theta(\mathbf{G}) \;\; \text{bits / dimension}.\label{eq:lattice_norment}
\end{align}
\fi
It was shown in~\cite{rogers1959lattice,rogers1964packing} that 
\iflongver
there is a sequence of lattices achieving Roger's bound
\else
there are lattices with
\fi
\begin{align}
\frac{1}{n}\log\Theta(\mathbf{G})\le \frac{\log n}{n}+(\log \sqrt{2\pi e})\frac{\log\log n}{n}+O\Big(\frac{1}{n}\Big),\label{eq:rogers}
\end{align}
and hence the redundancy w.r.t. maximum error for the optimal lattice quantizer is $O((\log n)/n)$.
This scaling order is precise for the optimal lattice quantizer since it was shown in \cite{coxeter1959covering} that the smallest possible covering density $\Theta(\mathbf{G})$ is at least $(1-o(1)) e^{-3/2} n$ 
as $n \to \infty$,
implying that $(1/n) \log \Theta(\mathbf{G})$ is at least in the order $(\log n)/n$.

We also study the \emph{mean squared error} (MSE).
For a random signal $\mathbf{X}$ uniform over a large set,
the MSE $\mathbb{E}[\Vert Q_{\mathcal{V}}(\mathbf{X})-\mathbf{X}\Vert^{2}]$
is close to the \emph{second moment} of the lattice \cite{zador1982asymptotic}
\begin{equation}
\frac{1}{\mu(\mathcal{V})}\int_{\mathcal{V}}\Vert\mathbf{x}\Vert^{2}\mathrm{d}\mathbf{x}.\label{eq:second_moment}
\end{equation}
The \emph{normalized second moment (NSM)}
\cite{zamir2014} is often studied:
\begin{align}
& G_n(\mathcal{V}):=\frac{\int_{\mathcal{V}}\Vert\mathbf{x}\Vert^{2}\mathrm{d}\mathbf{x}}{n (\mu (\mathcal{V}))^{1+2/n}}, \label{eq:nsm}
\end{align}
which is invariant under scaling ($G_n(\mathcal{V})=G_n(\gamma \mathcal{V})$
for $\gamma>0$). 
The Shannon redundancy w.r.t. MSE~\eqref{eq:red_mse_shannon} can be computed as
\begin{align}
\frac{1}{2} \log\big(2\pi e G_{n}(\mathcal{V})\big)  \;\; \text{bits / dimension},  \label{eq:lattice_mse}
\end{align}
which is also the high resolution limit of the redundancy of universal quantization~\cite{zamir1992universal}.
The Zador redundancy w.r.t. MSE~\eqref{eq:red_mse_zador} can be computed as
\begin{align}
\frac{1}{2}\log((n+2)G_{n}(\mathcal{V}))+\frac{1}{n}\log\kappa_{n}  \;\; \text{bits / dimension}. \label{eq:lattice_mse_zador}
\end{align}
The fact that \eqref{eq:lattice_mse_zador} is nonnegative, i.e., $G_{n}(\mathcal{V}) \ge 1/((n+2) \kappa_n^{2/n})$, is called Zador's lower bound~\cite{zador1982asymptotic}, and hence we call \eqref{eq:red_mse_zador} ``Zador redundancy''.

For achievability results, via a non-constructive proof, Zador~\cite{zador1982asymptotic} has shown that for every $n$, there exists a (non-lattice) vector quantizer achieving a Zador redundancy w.r.t. MSE~\eqref{eq:red_mse_zador} upper-bounded by (this is called \emph{Zador's upper bound})
\begin{align}
\frac{1}{2}\log\frac{(n+2)\Gamma(2/n+1)}{n},
    \label{eq:zador_ub}
\end{align}
and hence the Shannon redundancy w.r.t. MSE~\eqref{eq:red_mse_shannon} is $O((\log n)/n)$.
For lattice quantizers, it has been shown in \cite{zamir1996lattice,zamir2014} that Roger's bound~\eqref{eq:rogers} implies that there exist lattice quantizers with Shannon redundancy w.r.t. MSE being $O((\log n)/n)$.
Very recently, via a non-constructive proof, Ordentlich \cite[Theorem 2.5]{ordentlich2025voronoi} shows that there exists a lattice quantizer with $G_n(\mathcal{V}) \leq \Gamma(n/2 + 1)^{2/n}/(n\pi \sinc(2/n))$
where $\sinc(t):= \sin(\pi t)/(\pi t)$, which implies the following Zador redundancy w.r.t. MSE (see \cite[Equation (40)]{ordentlich2025voronoi})
\begin{align}
\frac{1}{2}\log\frac{n+2}{n\sinc(2/n)} \le \Big(\frac{1}{n}+\frac{4}{n^{2}}+\frac{8}{n^{3}}\Big)\log e\label{eq:or_bound}
\end{align}
for $n \geq 8$.\footnote{In Theorem 2.7 of \cite{ordentlich2025voronoi}, Ordentlich derives another upper bound on the NSM of lattices that converges to Zador's upper bound~\eqref{eq:zador_ub} as $n$ grows, thereby giving a positive indication that Gersho's conjecture \cite{gersho1979asymptotically} is correct.}
Overall, these results imply that the optimal redundancy w.r.t. maximum error for lattice quantizers is in the order $(\log n)/n$, the optimal Shannon redundancy w.r.t. MSE for lattice quantizers is in the order $(\log n)/n$, and the optimal Zador redundancy w.r.t. MSE for lattice quantizers is at most $O(1/n)$.

Given a basic cell $\mathcal{P}$, we can also construct the subtractively dithered
quantizer \cite{zamir1992universal}, which is a randomized quantizer
with random state $S=\mathbf{V}\sim\mathrm{Unif}(\mathcal{P})$, and
quantization function $Q(\mathbf{x},\mathbf{v})=Q_{\mathcal{P}}(\mathbf{x}-\mathbf{v})+\mathbf{v}$,
where $Q_{\mathcal{P}}$ is the deterministic lattice quantizer described
earlier. A useful property of dithered quantizer is that the quantization
error is always uniform over $\mathcal{P}$ and independent of the
input, i.e., $Q(\mathbf{x},\mathbf{V})-\mathbf{x}\sim\mathrm{Unif}(\mathcal{P})$
for every $\mathbf{x}\in\mathbb{R}^{n}$ (see \cite{kirac1996results,zamir2014}).
Therefore, when $\mathcal{P}$ is the Voronoi cell $\mathcal{V}$,
the MSE $\mathbb{E}[\Vert Q(\mathbf{X},\mathbf{V})-\mathbf{X}\Vert^{2}]$
exactly equals the second moment \eqref{eq:second_moment} regardless
of the distribution of $\mathbf{X}$. Nevertheless, the dithered quantizer
still has a normalized entropy $-\log|\det\mathbf{G}|$, and a maximum
error $\overline{\lambda}(\mathbf{G})$ when $\mathcal{P}$ is the
Voronoi cell $\mathcal{V}$. The dithered quantizer does not provide
any advantage in terms of normalized entropy and maximum error compared
to deterministic lattice quantizers. 

\myskip

\section{Rejection-Sampled Universal Quantizers}

\subsection{RSUQ with Error Uniform over a General Set}

We introduce a randomized quantizer with an error uniformly distributed
over a set $\mathcal{A}$ which is a subset of a basic cell, based
on applying rejection sampling on top of universal quantization \cite{zamir1992universal}.
Intuitively, we keep generating new dither signals until the error
falls in $\mathcal{A}$.

\medskip{}

\begin{defn}
\label{def:rej_samp_quant}Given a basic cell $\mathcal{P}$ of the
lattice $\mathbf{G}\mathbb{Z}^{n}$, and a subset $\mathcal{A}\subseteq\mathcal{P}$,
the \emph{rejection-sampled universal quantizer} (RSUQ) for $\mathcal{A}$
against $\mathcal{P}$ is the randomized quantizer $(P_{S},Q_{\mathcal{A},\mathcal{P}})$,
where $S=(\mathbf{V}_{i})_{i\in\mathbb{N}^{+}}$, $\mathbf{V}_{1},\mathbf{V}_{2},\ldots\stackrel{iid}{\sim}\mathrm{Unif}(\mathcal{P})$
are i.i.d. dither signals, and the quantization function
is
\begin{equation}
Q_{\mathcal{A},\mathcal{P}}(\mathbf{x},(\mathbf{v}_{i})_{i}):=Q_{\mathcal{P}}(\mathbf{x}-\mathbf{v}_{k})+\mathbf{v}_{k},\label{eq:QAP_def}
\end{equation}
\iflongver
where
\fi
\begin{equation}
k:=\min\big\{ i:\,Q_{\mathcal{P}}(\mathbf{x}-\mathbf{v}_{i})+\mathbf{v}_{i}-\mathbf{x}\in\mathcal{A}\big\},\label{eq:kstar}
\end{equation}
and $Q_{\mathcal{P}}$ is the lattice quantizer for basic cell
$\mathcal{P}$ (Section \ref{sec:lattice}).
\end{defn}
\medskip{}

Note that subtractive dithering \cite{ziv1985universal,zamir1992universal} is a special
case of RSUQ where $\mathcal{A}=\mathcal{P}$.\footnote{Technically, RSUQ requires a sequence of dithers $\mathbf{V}_{1},\mathbf{V}_{2},\ldots$,
whereas subtractive dithering only requires one dither, though when
$\mathcal{A}=\mathcal{P}$, only the first $\mathbf{V}_{1}$ is used.} Although RSUQ requires a sequence of dithers $\mathbf{V}_{1},\mathbf{V}_{2},\ldots$,
they can practically be generated by sharing a small random seed among
the encoder and the decoder, which can be used to generate the sequence
using a pseudorandom number generator (PRNG).\footnote{In this paper, we assume that the amount of shared randomness is unlimited. 
In practice, the encoder and the decoder can utilize a PRNG initialized with a shared random seed to approximately generate the sequence of dithers $\mathbf{V}_{1},\mathbf{V}_{2},\ldots$. 
Since the encoder and the decoder only need to share the random seed once, and reuse the same PRNG for all communication tasks, the communication cost of the random seed is inconsequential, and hence is omitted in all further theoretical analyses.}

To carry out the RSUQ in practice, the encoder and the decoder each
has a PRNG initialized with the same seed, so the outputs of the two
PRNGs will be the same. The encoder observes $\mathbf{X}\in\mathbb{R}^{n}$,
generates $\mathbf{V}_{1},\mathbf{V}_{2},\ldots\stackrel{iid}{\sim}\mathrm{Unif}(\mathcal{P})$
from the PRNG until iteration $K$ where $Q_{\mathcal{P}}(\mathbf{X}-\mathbf{V}_{K})+\mathbf{V}_{K}-\mathbf{X}\in\mathcal{A}$,
computes $\mathbf{M}:=Q_{\mathcal{P}}(\mathbf{X}-\mathbf{V}_{K})\in\mathbf{G}\mathbb{Z}^{n}$,
and encodes and transmits $(K,\mathbf{M})$. 
\ifshortver
If we know that $\mathbf{X}\in \mathcal{X}$, then we can compress $K$ using the optimal prefix-free code for $\mathrm{Geom}(\mu(\mathcal{A})/\mu(\mathcal{P}))$ \cite{Golomb1966enc,Gallager1975Geom}, and compress $\mathbf{M} \in \mathcal{M} := (\mathcal{X}+\mathcal{A}-\mathcal{P}) \cap \mathbf{G}\mathbb{Z}^{n}$ using $\lceil \log |\mathcal{M}|\rceil$ bits. 
\fi
The decoder observes
$(K,\mathbf{M})$, generates $\mathbf{V}_{1},\ldots,\mathbf{V}_{K}\stackrel{iid}{\sim}\mathrm{Unif}(\mathcal{P})$
from the PRNG, and outputs $\mathbf{Y}=\mathbf{M}+\mathbf{V}_{K}$. 
The encoding and decoding operations of RSUQ are given in  Algorithms~\ref{alg:RSUQ_encode} and~\ref{alg:RSUQ_decode}, respectively.

\begin{algorithm}[ht]
\textbf{$\;\;\;\;$Input:} vector  $\mathbf{x}\in\mathbb{R}^n$, full-rank generator matrix $\mathbf{G} \in \mathbb{R}^{n\times n}$, basic cell $\mathcal{P}$ of $\mathbf{G}\mathbb{Z}^n$, subset $\mathcal{A}\subseteq\mathcal{P}$, random number generator (RNG) $\mathfrak{B}$ 

\textbf{$\;\;\;\;$Output:} description $(K,
\mathbf{M}) \in \mathbb{N}^{+} \times \mathbf{G}\mathbb{Z}^n$ 

\smallskip{}

\begin{algorithmic}[1]
\For{$i=1, 2, \ldots$} 
\State \label{step:RS1} Sample 
$\mathbf{V}_{i} \sim\mathrm{Unif}(\mathcal{P})$ using RNG $\mathfrak{B}$
\State $\mathbf{M} \leftarrow Q_{\mathcal{P}}(\mathbf{x}-\mathbf{V}_{i})\in\mathbf{G}\mathbb{Z}^{n}$
\If{$\mathbf{M}+\mathbf{V}_{i}-\mathbf{x}\in \mathcal{A}$} 
\State \Return $(i ,\mathbf{M})$
\EndIf
\EndFor
\end{algorithmic}

\caption{\label{alg:RSUQ_encode}$\textsc{RSUQ\_Encode}(\mathbf{x},\mathbf{G},\mathcal{P},\mathcal{A},\mathfrak{B})$}
\end{algorithm}

\begin{algorithm}[ht]
\textbf{$\;\;\;\;$Input:} description $(K,\mathbf{M}) \in \mathbb{N}^{+}\times \mathbf{G}\mathbb{Z}^n$, full-rank generator matrix $\mathbf{G} \in \mathbb{R}^{n\times n}$, basic cell $\mathcal{P}$ of $\mathbf{G}\mathbb{Z}^n$, RNG $\mathfrak{B}$

\textbf{$\;\;\;\;$Output:} $\mathbf{Y} \in \mathbb{R}^{n} $ 

\smallskip{}

\begin{algorithmic}[1]
\State Sample 
$\mathbf{V}_{1},\ldots \mathbf{V}_{K} \stackrel{iid}{\sim} \mathrm{Unif}(\mathcal{P})$ using RNG $\mathfrak{B}$  
\State \Return $\mathbf{M}+\mathbf{V}_{K}$ 
\end{algorithmic}

\caption{\label{alg:RSUQ_decode}$\textsc{RSUQ\_Decode}((K,\mathbf{M}),\mathbf{G},\mathcal{P},\mathfrak{B})$}
\end{algorithm}

We can check that the error is uniform over $\mathcal{A}$. 
\ifshortver
The proof is given in 
\cite{ling2024rejectionsampled}.
\else
The proof is given in Appendix~\ref{subsec:pf_error_dist_unif}.
\fi

\medskip{}

\begin{prop} \label{prop:error_dist_unif}
For any $\mathbf{x} \in \mathbb{R}^n$, the quantization error $\mathbf{Z} := Q_{\mathcal{A},\mathcal{P}}(\mathbf{x},S) - \mathbf{x}$ of the RSUQ $(P_S, Q_{\mathcal{A},\mathcal{P}})$ (where $S \sim P_S$) 
\ifshortver
follows the uniform distribution $\mathbf{Z} \sim \mathrm{Unif}(\mathcal{A})$.
\else
follows the uniform distribution over the set $\mathcal{A}$, i.e., $\mathbf{Z} \sim \mathrm{Unif}(\mathcal{A})$.
\fi
\end{prop}
\medskip{}

We now bound the normalized entropy of RSUQ. 
The proof is given in Appendix~\ref{subsec:pf_rsuq_ent}.
\myskip
\iflongver
\begin{thm}
\label{thm:rsuq_ent}Consider the RSUQ
$Q_{\mathcal{A},\mathcal{P}}$ for $\mathcal{A}$ against $\mathcal{P}$.
For a random input $\mathbf{X}$ satisfying $\mathbf{X}\in\mathcal{X}$
almost surely for some $\mathcal{X}\subseteq\mathbb{R}^{n}$ with
$\mu(\mathcal{X})<\infty$, independent of $S\sim P_{S}$, the conditional
entropy of the quantizer satisfies
\begin{align}
 & H(Q_{\mathcal{A},\mathcal{P}}(\mathbf{X},S)\,|\,S)\nonumber \\
 & \le\log\mu(\mathcal{X}+\mathcal{A}+\mathcal{P}-\mathcal{P})-\log\mu(\mathcal{A})-\frac{1-p}{p}\log(1-p)\nonumber \\
 & \le\log\mu(\mathcal{X}+\mathcal{A}+\mathcal{P}-\mathcal{P})-\log\mu(\mathcal{A})+\log e,\label{eq:HqXS}
\end{align}
where $p:=\mu(\mathcal{A})/\mu(\mathcal{P})$. Hence, the normalized
entropy 
$\overline{H}(Q_{\mathcal{A},\mathcal{P}})$ is upper-bounded by
\begin{align*}
 & -\log\mu(\mathcal{A})-\frac{1-p}{p}\log(1-p) \; \le \; -\log\mu(\mathcal{A})+\log e.
\end{align*}
\end{thm}
\else
\begin{thm}
\label{thm:rsuq_ent}Let $p:=\mu(\mathcal{A})/\mu(\mathcal{P})$. The normalized
entropy of RSUQ
$\overline{H}(Q_{\mathcal{A},\mathcal{P}})$ is upper-bounded by
\begin{align*}
 & -\log\mu(\mathcal{A})-\frac{1-p}{p}\log(1-p) \; \le \; -\log\mu(\mathcal{A})+\log e.
\end{align*}
\end{thm}
\fi
\myskip

\iflongver
The bound~\eqref{eq:HqXS} works best when $\mathbf{X}$ is uniformly distributed. We can obtain tighter bounds on the conditional entropy $H(Q_{\mathcal{A},\mathcal{P}}(\mathbf{X},S)\,|\,S)$ for nonuniform $\mathbf{X}$ in terms of its differential entropy, in a similar manner as \eqref{eqn:lattice_diffent}, which will be elaborated in Section~\ref{sec:nonunif_input}.

The proof of Theorem~\ref{thm:rsuq_ent} suggests that, operationally, 
if we know that $\mathbf{X}\in \mathcal{X}$, then we should compress $K$ using the optimal prefix-free code for $\mathrm{Geom}(p)$ \cite{Golomb1966enc,Gallager1975Geom}, and then compress $\mathbf{M}:=Q_{\mathcal{P}}(\mathbf{X}-\mathbf{V}_{K}) \in \mathcal{M} := (\mathcal{X}+\mathcal{A}-\mathcal{P}) \cap \mathbf{G}\mathbb{Z}^{n}$ using $\lceil \log |\mathcal{M}|\rceil$ number of bits. 
This encoding scheme has an advantage that the construction does not depend on the distribution of $\mathbf{X}$, and has the same expected length regardless of the distribution of $\mathbf{X}$.
\fi

\begin{figure*}[ht]
\begin{centering}
\includegraphics[scale=0.40]{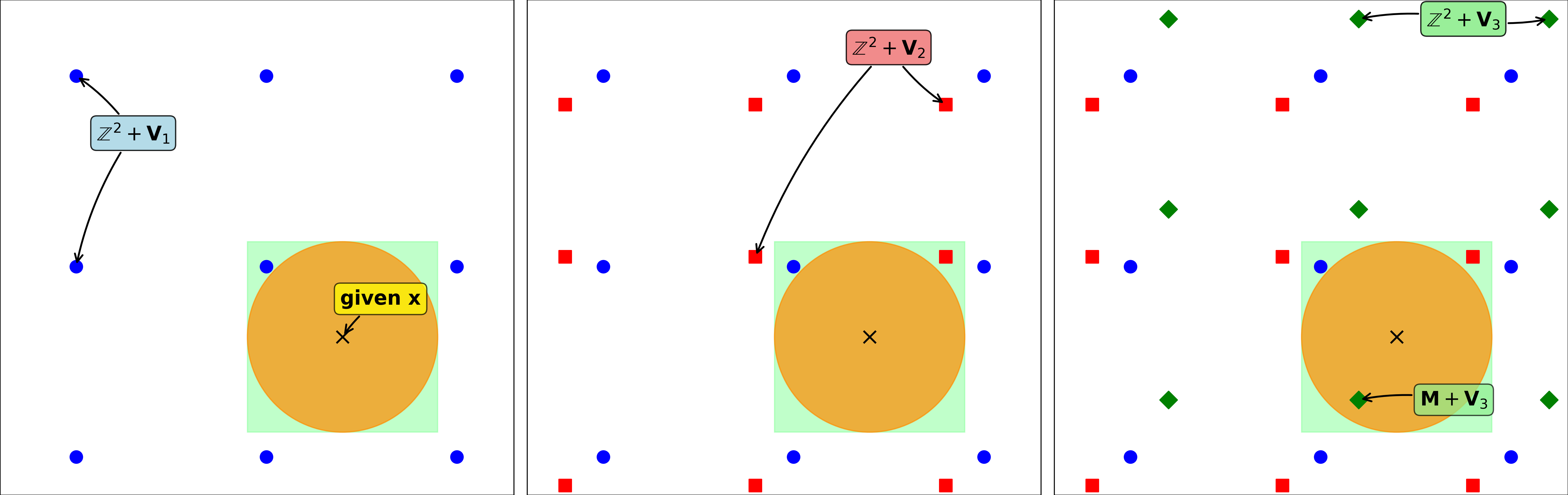} 
\par\end{centering}
\caption{\label{fig:RSUQ} RSUQ with $\mathbf{G} = \mathbf{I}_2$ (the $2 \times 2$ identity matrix), yielding the two-dimensional integer lattice $\mathbb{Z}^2$ with the basic cell $\mathcal{P}=\left(-1/2,1/2\right]^2$ (the Voronoi cell), and $\mathcal{A} = (1/2)\cdot B^2$ (a two-dimensional ball with radius $1/2$).
Fix some $\mathbf{x} \in \mathbb{R}^2$.
For $i=1,2,\ldots$, generate $\mathbf{V}_i \in \left(-1/2,1/2\right]^2$ and find the unique $\mathbf{M} \in \mathbb{Z}^2$ such that $\mathbf{M}+\mathbf{V}_i \in (\mathbb{Z}^2+\mathbf{V}_i) \cap (\mathbf{x}+\left(-1/2,1/2\right]^2)$.
The first and second iterations (left and middle figures) are rejected because $\mathbf{M}+\mathbf{V}_i-\mathbf{x} \notin (1/2)\cdot B^2$, for $i = 1, 2$, and the third iteration (right figure) is accepted because $\mathbf{M}+\mathbf{V}_3-\mathbf{x} \in (1/2)\cdot B^2$.
}
\vspace{-3pt}
\end{figure*}

\subsection{RSUQ with Error Uniform over an $n$-Ball}

In particular, we are interested in the RSUQ for the ball $rB^{n}$. 
In this case, the MSE is exactly $nr^{2}/(n+2)$ regardless of the distribution of $\mathbf{X}$, unlike deterministic lattice quantizers where the MSE is the second moment~\eqref{eq:second_moment} only for $\mathbf{X}$ that is sufficiently spread out.
We have the
following corollary of Theorem~\ref{thm:rsuq_ent}.
\ifshortver
The proof is in \cite{ling2024rejectionsampled}.
\else
The proof is in Appendix~\ref{subsec:pf_rsuq_ball}.
\fi

\myskip

\begin{cor}\label{cor:rsuq_ball}
Fix any $r>0$ and generator matrix $\mathbf{G}\in\mathbb{R}^{n\times n}$.
There exists an RSUQ for the ball $rB^{n}$ against the Voronoi cell
$\mathcal{V}$ of some lattice, with normalized
entropy 
\begin{align}
 \overline{H}(Q) & \le -n\log r-\log\kappa_{n}-\frac{1-\delta(\mathbf{G})}{\delta(\mathbf{G})}\log(1\! -\! \delta(\mathbf{G}))\label{eq:rsuq_ent_ub}\\
 & \le-n\log r-\log\kappa_{n}+\log e\label{eq:rsuq_ent_ub_ar}
\end{align}
where $\delta(\mathbf{G})$ is the packing density \eqref{eq:density}. Since this RSUQ has an error distribution $\mathrm{Unif}(rB^n)$, it has a maximum error $r$ and MSE $\mathbb{E}[\Vert Q(\mathbf{X},S)-\mathbf{X}\Vert^{2}]= nr^{2}/(n+2)$.
\end{cor}

\myskip

While the upper bound \eqref{eq:rsuq_ent_ub} depends on the packing density $\delta(\mathbf{G})$ of the lattice, the upper bound \eqref{eq:rsuq_ent_ub_ar} does not depend on the lattice. Therefore, the tradeoff between maximum error (or MSE) and normalized entropy of RSUQ is not sensitive to the choice of the lattice, in stark contrast to conventional lattice quantizers where the lattice must be designed carefully.
Intuitively, this is due to the fact that the performance of a lattice quantizer depends on the shape of its basic cell (i.e., how close it is to a ball), which highly depends on the choice of lattice. In comparison, the rejection sampling in RSUQ always turns the error distribution into a uniform distribution over a ball, regardless of the shape of the basic cell of the lattice it uses.
Moreover, for RSUQ, we can choose a lattice with a fast and simple quantization algorithm, e.g., the integer lattice $\mathbb{Z}^n$ or a root lattice~\cite{conway1982fast}. In contrast, for a good conventional lattice quantizer, a fast algorithm is often not easy to construct (e.g., \cite{vardy1993maximum}). 
Refer to Section~\ref{sec:timeRSUQ} for further discussion on time complexity.

Comparing the upper bound in \eqref{eq:rsuq_ent_ub_ar} and the lower bounds in Proposition \ref{prop:lb}, we can see that the RSUQ has a 
redundancy w.r.t. maximum error~\eqref{eq:red_maxerror} (per dimension) at most 
\begin{align}
\frac{\log e}{n} \;\; \text{bits / dimension}, \label{eq:rsuq_maxerror_red}
\end{align}
significantly smaller than the $O((\log n)/n)$ redundancy for the optimal lattice quantizer in~\eqref{eq:rogers}.
It also has a Zador redundancy w.r.t. MSE~\eqref{eq:red_mse_shannon}
\begin{align}
\frac{\log e}{n} \;\; \text{bits / dimension}, \label{eq:rsuq_mse_red_zador}
\end{align} 
which is tighter than Ordentlich's bound $(n^{-1}+4n^{-2}+8n^{-3})\log e$ for lattice quantizers in \eqref{eq:or_bound} since the terms $(4n^{-2}+8n^{-3})\log e$ are absent in our bound, but is not as tight as Zador's upper bound \eqref{eq:zador_ub} which is non-constructive.
It has a Shannon redundancy w.r.t. MSE~\eqref{eq:red_mse_shannon} at most 
\begin{align}
&\frac{1}{2}\log\frac{2\pi e}{n+2}+\frac{1}{n}\log\frac{e}{\kappa_{n}} \;\le\; \frac{\log n}{2n} + \frac{1.4}{n} \;\; \text{bits / dimension}, \label{eq:rsuq_mse_red_ar}
\end{align}
(see Appendix~\ref{subsec:pf_mse_gap}), 
the same scaling as the optimal lattice quantizer~\eqref{eq:zador_ub}, 
but is significantly easier to construct due to the insensitivity to the choice of the lattice.

We also remark that if a deterministic quantizer is desired, we can fix the dithers $\mathbf{V}_1,\mathbf{V}_2,\ldots$ for the RSUQ to give a good deterministic entropy-constrained vector quantizer. One can show that there exists a fixed choice of $\mathbf{V}_1,\mathbf{V}_2,\ldots$ that gives a deterministic quantizer with redundancy w.r.t. maximum error at most $(\log e)/n$, and there exists a fixed choice of $\mathbf{V}_1,\mathbf{V}_2,\ldots$ that gives a deterministic quantizer with Zador redundancy w.r.t. MSE at most $(\log e)/n$.\footnote{To show this, let $\mathbf{V}=(\mathbf{V}_1,\mathbf{V}_2,\ldots)$, $\overline{H}(Q|\mathbf{v})$ be the normalized entropy of RSUQ when we fix $\mathbf{V}=\mathbf{v}$, and $D(\mathbf{v})$ be the MSE when we fix $\mathbf{V}=\mathbf{v}$. Since $\mathbb{E}[\overline{H}(Q|\mathbf{V})]=\overline{H}(Q)$, there must exist a fixed $\mathbf{v}$ with $\overline{H}(Q|\mathbf{v}) \leq \overline{H}(Q)$, and hence has a redundancy w.r.t. maximum error at most $(\log e)/n$. Also, since $\mathbb{E}[n^{-1}\overline{H}(Q|\mathbf{V})+(1/2)\log D(\mathbf{V})]\le n^{-1}\overline{H}(Q)+(1/2)\log D$ by Jensen's inequality, there must exist a fixed $\mathbf{v}$ with Zador redundancy w.r.t. MSE \eqref{eq:red_mse_zador} at most $(\log e)/n$. } 
Nevertheless, once $\mathbf{V}_1,\mathbf{V}_2,\ldots$ is fixed, the error distribution will no longer be uniform over a ball.

\begin{figure*}
\begin{centering}
\includegraphics[scale=0.47]{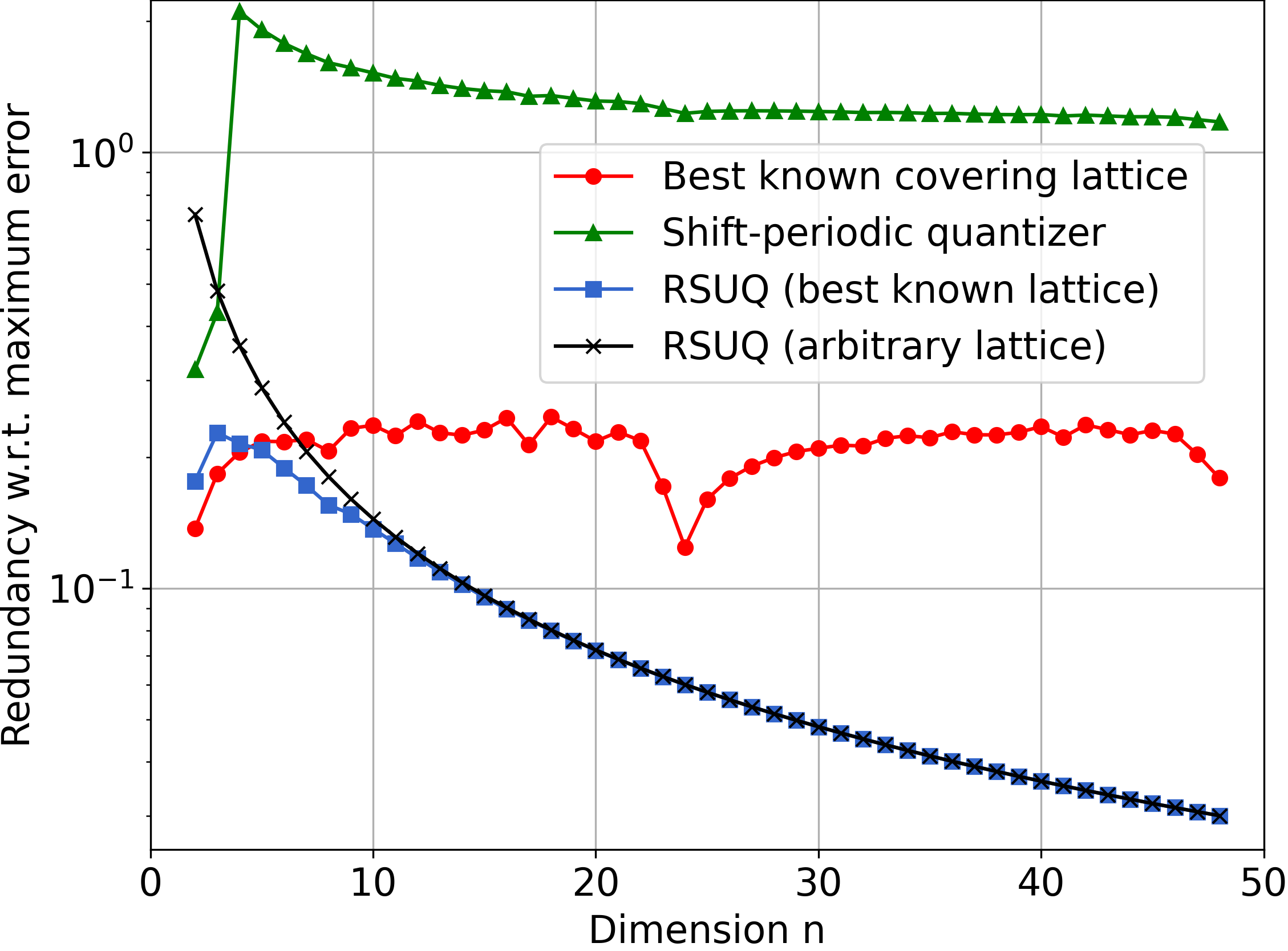} \includegraphics[scale=0.47]{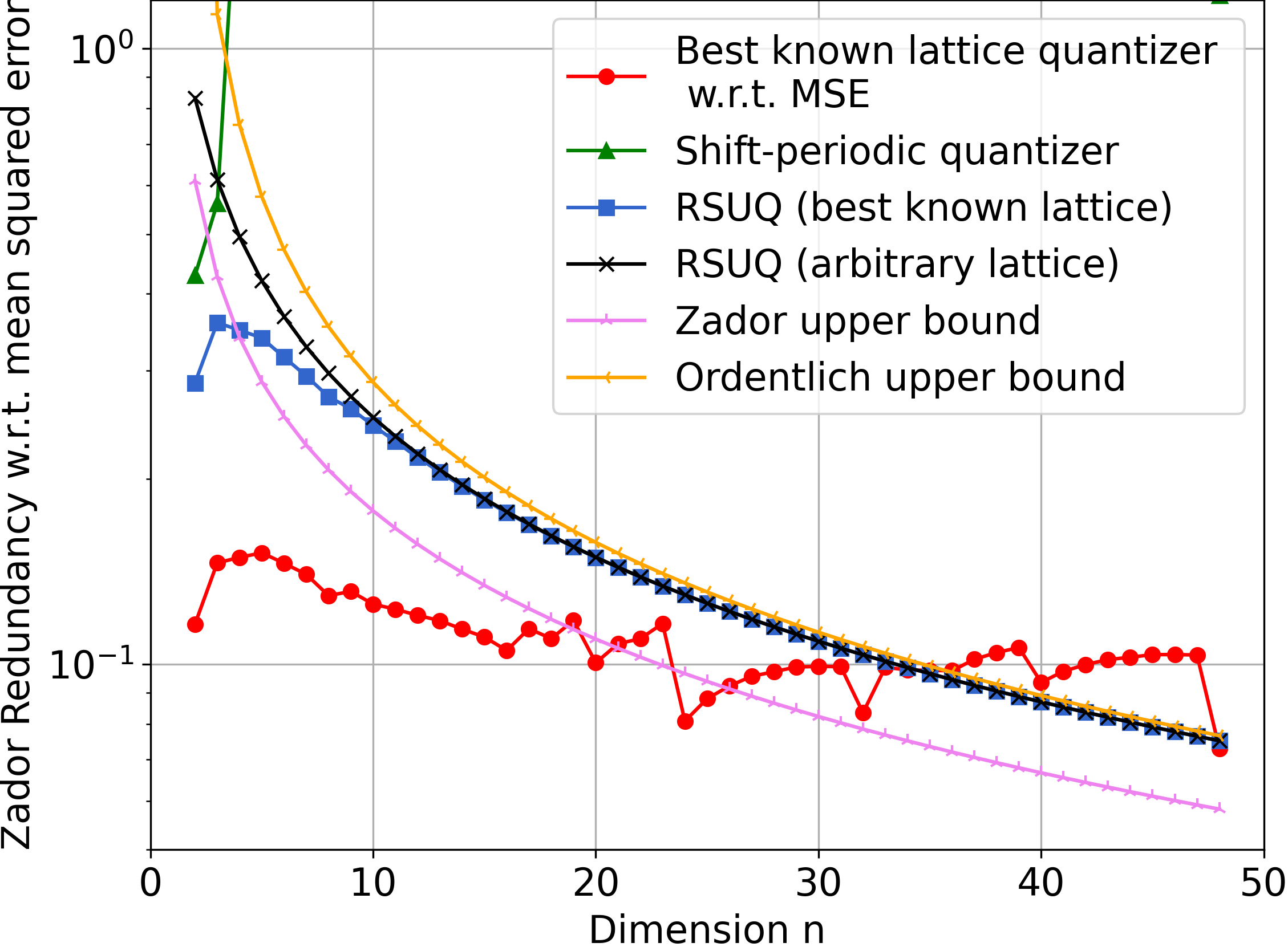}
\par\end{centering}
\caption{\label{fig:compare_lattice}Log-scale plot of the redundancy w.r.t. maximum error~\eqref{eq:red_maxerror} (left figure) and the Zador redundancy w.r.t. MSE~\eqref{eq:red_mse_zador} (right figure) of the 
lattice quantizer with the best known covering radius $\overline{\lambda}(\mathbf{G})$ on the left, and with the best known NSM  $G_n(\mathcal{V})$ on the right
(red line),
the shift-periodic quantizer (green line),
the RSUQ with error distribution $\mathrm{Unif}(B^{n})$ constructed
using the best known sphere packing lattice  (blue line), 
the RSUQ constructed using an arbitrary lattice (black line), Zador's upper bound \eqref{eq:zador_ub} (violet line), and Ordentlich's upper bound \eqref{eq:or_bound} (orange line).}
\vspace{-3pt}
\end{figure*}

Figure \ref{fig:compare_lattice} (left figure) plots the following for dimension $n=2,\ldots,48$:
\begin{itemize}
\item The redundancy w.r.t. maximum error~\eqref{eq:red_maxerror} 
of lattice quantizer, using the best known sphere covering lattice (with $\Theta(\mathbf{G})$
given in \cite{sikiric2008generalization}), computed using \eqref{eq:lattice_norment}.\footnote{Note that \cite{sikiric2008generalization} only gave $\Theta(\mathbf{G})$ up to $n=24$. For $n=25,\ldots,48$, $\Theta(\mathbf{G})$ is computed using the optimal product lattice among the lattices in \cite{sikiric2008generalization} which, to the best of the authors' knowledge, are the best known $\Theta(\mathbf{G})$.}
\item The redundancy of shift-periodic quantizers \cite{ling2023vector}, which can also ensure an error distribution $\mathrm{Unif}(B^{n})$.\footnote{For $n=2,3$, we use the bounds computed in the examples in \cite{ling2023vector}. For $n\ge 4$, we use the general bound in \cite[Thm 11 part 1]{ling2023vector}.}
\item The upper bound on the redundancy w.r.t. maximum error 
of RSUQ with error distribution $\mathrm{Unif}(B^{n})$ computed using \eqref{eq:rsuq_ent_ub} and the
best known lattice for sphere packing (with $\delta(\mathbf{G})$
given in \cite{cohn2024sphere}).
\item The general upper bound $(\log e)/n$ on the redundancy w.r.t. maximum error 
of RSUQ with error distribution $\mathrm{Unif}(B^{n})$ by \eqref{eq:rsuq_ent_ub_ar}, which applies
to any lattice.
\end{itemize}
Figure \ref{fig:compare_lattice} (left figure) highlights two advantages of
RSUQ:
\begin{itemize}
\item RSUQ (with the best lattice for sphere packing) achieves a smaller redundancy
(and hence a smaller normalized entropy for the same maximum
error)
compared to the best known lattice quantizer for $n\in\{5,\ldots,48\}$.%
\footnote{One objection to this comparison is that if we actually encode
the description into bits, then there is an at most $1$ bit penalty
incurred by Huffman coding \cite{huffman1952method} (or a $1/n$ bit penalty on the redundancy). Note that both
the lattice quantizer and RSUQ are subject to this penalty. Moreover,
even if we impose this penalty on RSUQ and do not impose
this penalty on lattice quantizers, RSUQ still outperforms
lattice quantizers for $n\in\{11,\ldots,48\}$.
}
\item Even with an arbitrary lattice, RSUQ achieves a smaller  redundancy compared to the best known lattice quantizer for $n\in\{7,\ldots,48\}$.
\iflongver
RSUQ is insensitive to the choice of lattice, since the
gap between the (per dimension) redundancy implied by \eqref{eq:rsuq_ent_ub} (which depends
on the lattice) and the redundancy implied by \eqref{eq:rsuq_ent_ub_ar} (which holds for every
lattice) is $O(1/n)$ as $n\to\infty$. 
\fi
This makes the construction
of RSUQ simpler than that of lattice quantizer, where the performance
of the latter is highly sensitive to the construction of the lattice,
which is often not easy.
\end{itemize}

We also carry out the same comparison for mean squared error instead
of maximum error in Figure \ref{fig:compare_lattice} (right figure), i.e., 
we consider the Zador redundancy w.r.t. MSE~\eqref{eq:red_mse_zador} instead of maximum error.
\iflongver
We consider the dimensions $n=2,\ldots,48$. 
\fi
We plot the following:
\begin{itemize}
\item The Zador redundancy of the best known
lattice quantizers, obtained from the formula \eqref{eq:lattice_mse_zador},
with the best known $G_n(\mathcal{V})$ from \cite{agrell2023best,Kudryashov2010LowComplexLattice,Lyu2022BetterLatticeQuan,Agrell2024gluedLat,Agrell2025OptLatticeQuan}. Refer to Appendix~\ref{subsec:NSM_detailed_plot} for a detailed plot where each of \cite{agrell2023best,Kudryashov2010LowComplexLattice,Lyu2022BetterLatticeQuan,Agrell2024gluedLat,Agrell2025OptLatticeQuan} is plotted separately.
\item Zador's upper bound~\eqref{eq:zador_ub},
which is a non-constructive upper bound on the best (non-lattice) vector quantizer. 
\item Ordentlich's upper bound (left-hand side of~\eqref{eq:or_bound}), which is a non-constructive upper bound on the best lattice quantizer. 
\item The Zador redundancy of shift-periodic quantizers \cite{ling2023vector}.
\item The upper bound on the Zador redundancy of RSUQ with the best known sphere packing lattice, computed using
\eqref{eq:red_mse_zador} and \eqref{eq:rsuq_ent_ub} in Corollary~\ref{cor:rsuq_ball}.
\item The $(\log e)/n$ upper bound \eqref{eq:rsuq_mse_red_zador} on the Zador redundancy of RSUQ with arbitrary lattice.
\end{itemize}

RSUQ achieves a smaller redundancy compared to the best known lattice quantizer for $n\in\{35,\ldots,47\}$. 
The right panel of Figure~\ref{fig:compare_lattice} shows that the RSUQ likely does not improve
on the best possible 
vector quantizer (Zador's upper bound \eqref{eq:zador_ub}, which is non-constructive), but it improves on the best known explicit constructions of
lattice quantizers with respect to MSE for $n \in \{35,\ldots,47\}$, and also on Ordentlich's upper bound~\eqref{eq:or_bound}. 
This, combined with the fact that RSUQ can be constructed using simple lattices like the integer lattice $\mathbb{Z}^n$, makes RSUQ a competitive choice compared to conventional lattice quantizers, which are highly nontrivial to construct explicitly for high dimensions.

\subsection{Time Complexity of RSUQ} \label{sec:timeRSUQ}

We now discuss the running time complexity of RSUQ.  
The number of steps needed for rejection sampling is a geometric random variable $K \sim \mathrm{Geom}(\delta(\mathbf{G}))$ with expectation $1/\delta(\mathbf{G})$, where $\delta(\mathbf{G})$ is the packing density of the lattice $\mathbf{G}\mathbb{Z}^n$.
RSUQ is a randomized algorithm with an overall expected time complexity $O(\mathrm{LDP}(\mathbf{G})/\delta(\mathbf{G}))$, where $\mathrm{LDP}(\mathbf{G})$ is the complexity of the lattice decoding problem for $\mathbf{G}$. 
With probability $1-\epsilon$ (where $0<\epsilon<1$), RSUQ will terminate within $K \le \lceil \ln(1/\epsilon) / \delta(\mathbf{G}) \rceil$ steps, and hence RSUQ terminates within a time $O(\mathrm{LDP}(\mathbf{G}) \ln(1/\epsilon)/\delta(\mathbf{G}))$ with probability $1-\epsilon$.
Therefore, choosing a lattice with large $\delta(\mathbf{G})$ can improve the running time.

Since the order of growth of $1/\delta(\mathbf{G})$ is at least $2^{\Omega(n)}$ \cite{conway2013sphere}, the complexity of RSUQ is at least exponential in $n$, though this is not a disadvantage of RSUQ compared with conventional lattice quantizers. 
If we simply fix $\mathbf{G}$ to be the integer lattice $\mathbb{Z}^n$ with $\mathrm{LDP}(\mathbf{G})=O(n)$,
the complexity of RSUQ is $2^{(n/2)\log n + O(n)}$, which is generally better than the $2^{(n/2)\log (\gamma n)+O(n)}$ complexity for conventional lattice quantizers (for general lattices with equal successive minima, where $\gamma \in [1,n]$ is the coding gain) \cite[Theorem 2]{banihashemi1998complexity}.\footnote{ This comparison holds only for generic lattices, not for those with known lower-complexity algorithms. For example, Kudryashov and Yurkov \cite{Kudryashov2010LowComplexLattice} proposed a class of lattices with encoding complexity linear in the lattice dimension $n$. 
Only the dimensions $n = 12, 16, 20, 24, 32, 48, 96$ have been evaluated in \cite{Kudryashov2010LowComplexLattice}. Among these $n$'s, our scheme attains an MSE smaller than \cite{Kudryashov2010LowComplexLattice} for $n=96$, but the MSE is larger for $n = 12, 16, 20, 24, 32, 48$.} 
Therefore, RSUQ applied on the integer lattice $\mathbb{Z}^n$ can outperform conventional lattice quantizers applied on general lattices, in the sense that RSUQ attains a smaller redundancy w.r.t. maximum error, while having a running time complexity at least as good as conventional lattice quantizers, and also having a significantly simpler algorithm. 
More discussions on the time complexity of our scheme applied to channel simulation will be given in Remark~\ref{remark:time}.

\myskip

\section{RSUQ with Nonuniform Input Distribution and Universal Quantization}\label{sec:nonunif_input}

\iflongver
We then study the conditional entropy of the RSUQ for nonuniformly distributed $\mathbf{X}$. Recall that the normalized entropy $\overline{H}(Q)$ (Definition \ref{def:norment}) describes the conditional entropy $H(Q(\mathbf{X},S)|S)$ of the randomized quantizer for uniform input, in the sense that $H(Q(\mathbf{X},S)|S) = \overline{H}(Q)+h(\mathbf{X}) + o(1)$ when $\mathbf{X} \sim \mathrm{Unif}(\tau B^n)$ is uniformly distributed over a large ball as $\tau \to \infty$, where $h(\mathbf{X})$ is the differential entropy (we have $h(\mathbf{X}) = \log \mu(\tau B^n)$ for $\mathbf{X} \sim \mathrm{Unif}(\tau B^n)$). We show that for RSUQ, this relation holds for nonuniformly distributed $\mathbf{X}$ as well (see \eqref{eqn:condent_diffent}). The normalized entropy $\overline{H}(Q)$ can describe the conditional entropy of RSUQ for uniformly or nonuniformly distributed $\mathbf{X}$ as long as $\mathbf{X}$ is sufficiently spread out (i.e., the probability density function $f_{\mathbf{X}}(\mathbf{x})$ is sufficiently smooth at most $\mathbf{x}$), which will be shown in Theorem \ref{thm:rsuq_ent_gen}. 
This can be seen as an analogue to the result on conventional lattice quantizers in \eqref{eqn:lattice_diffent}.
The proof is given in Appendix~\ref{subsec:pf_rsqu_ent_gen}.

\medskip{}

\begin{thm} \label{thm:rsuq_ent_gen}
Consider the RSUQ $Q_{\mathcal{A}, \mathcal{P}}$ for $\mathcal{A}$ against the basic cell $\mathcal{P}$ of the lattice $ \mathbf{G}\mathbb{Z}^n$.
Fix any random input $\mathbf{X} \in \mathbb{R}^n$ with a probability density function $f_{\mathbf{X}}$ supported over $\mathcal{X} \subseteq \mathbb{R}^n$. We have
\begin{align*}
\overline{H}(Q_{\mathcal{A}, \mathcal{P}}) + h(\mathbf{X}) &\le H(Q_{\mathcal{A}, \mathcal{P}}(\mathbf{X},S)|S) \le \overline{H}(Q_{\mathcal{A}, \mathcal{P}}) + h(\mathbf{X}) + \overline{\delta}_{\epsilon},    
\end{align*}
where
\begin{align} \label{eq:bar_delta_eps}
\overline{\delta}_{\epsilon} &:= \int_{\mathcal{X}} \delta_{\epsilon}(\mathbf{x})f_{\mathbf{X}}(\mathbf{x}) \mathrm{d}\mathbf{x} \nonumber+(\log e) \Big(\sup_{\mathbf{x} \in \mathcal{X}}f_{\mathbf{X}}(\mathbf{x}) \Big) \mu \big((\mathcal{X} + \epsilon B^n) \backslash \mathcal{X} \big),    
\end{align}
\begin{equation}
    \delta_{\epsilon}(\mathbf{x}):=\sup_{\mathbf{x}' \in \mathcal{X} :\,\Vert\mathbf{x}-\mathbf{x}'\Vert\le \epsilon}\log \frac{f_{\mathbf{X}}(\mathbf{x})}{f_{\mathbf{X}}(\mathbf{x}')},
\end{equation}
and $\epsilon := \mathrm{diam}(\mathcal{A})$ is the diameter of $\mathcal{A}$.\footnote{We regard the term $(\log e) (\sup_{\mathbf{x} \in \mathcal{X}}f_{\mathbf{X}}(\mathbf{x}) ) \mu ((\mathcal{X} + \epsilon B^n) \backslash \mathcal{X})$ to be $0$ if $\mathcal{X}=\mathbb{R}^n$, regardless of whether $\sup_{\mathbf{x} \in \mathbb{R}^n}f_{\mathbf{X}}(\mathbf{x})$ is finite.} Combining this with Theorem~\ref{thm:rsuq_ent} gives
\[
H(Q_{\mathcal{A}, \mathcal{P}}(\mathbf{X},S)|S) \le h(\mathbf{X}) - \log \mu(\mathcal{A}) + \log e + \overline{\delta}_{\epsilon}.
\]
\end{thm}
\myskip

As a corollary, if we consider a scaled version $Q_{\alpha\mathcal{A}, \alpha\mathcal{P}}$ of the RSUQ, where the sets $\mathcal{A},\mathcal{P}$ and the lattice are scaled by $\alpha > 0$, then $H(Q(\mathbf{X},S)|S) = \overline{H}(Q)+h(\mathbf{X}) + o(1)$ holds in the high resolution limit where $\alpha \to 0$. 
This is analogous to the high resolution limit of conventional lattice quantizers in \eqref{eqn:lattice_diffent}.

\medskip{}

\begin{cor} \label{cor:rsuq_ent_gen_eps}
Consider the RSUQ $Q_{\alpha\mathcal{A}, \alpha\mathcal{P}}$ for $\alpha\mathcal{A}$ against the basic cell $\alpha\mathcal{P}$ of the lattice $\alpha \mathbf{G}\mathbb{Z}^n$, where $\alpha > 0$.
Fix any random input $\mathbf{X} \in \mathbb{R}^n$ with a probability density function $f_{\mathbf{X}}$ satisfying $\lim_{\epsilon \to 0} \overline{\delta}_{\epsilon} = 0$.
Then
\begin{equation}
    \lim_{\alpha \rightarrow 0}\big(H(Q_{\alpha\mathcal{A}, \alpha\mathcal{P}}(\mathbf{X},S)|S)- \overline{H}(Q_{\alpha\mathcal{A}, \alpha\mathcal{P}})\big) = h(\mathbf{X}).
\end{equation}
Combining this with Theorem~\ref{thm:rsuq_ent} gives
\begin{equation}
    \underset{\alpha \rightarrow 0}{\mathrm{limsup}}\, \big(H(Q_{\alpha\mathcal{A}, \alpha\mathcal{P}}(\mathbf{X},S)|S) + \log \mu(\alpha \mathcal{A})\big) \le  h(\mathbf{X}) + \log e.
\end{equation}
\end{cor}
\begin{IEEEproof}
By letting $\alpha \to 0$, and so $\epsilon \to 0$ since $\epsilon = \mathrm{diam} (\alpha \mathcal{A})$, we have $\lim_{\epsilon \to 0} \overline{\delta}_{\epsilon} = 0$. The result follows from Theorem~\ref{thm:rsuq_ent_gen}.
\end{IEEEproof}

\medskip{}

Corollary \ref{cor:rsuq_ent_gen_eps} requires the distribution of $\mathbf{X}$ to satisfy $\lim_{\epsilon \to 0} \overline{\delta}_{\epsilon} = 0$.
We now state some sufficient conditions for $\lim_{\epsilon \to 0} \overline{\delta}_{\epsilon} = 0$ to hold. The proof is given in Appendix \ref{subsec:pf_lim_deltabar}.

\medskip{}

\begin{prop}\label{prop:lim_deltabar}
The condition $\lim_{\epsilon \to 0} \overline{\delta}_{\epsilon} = 0$ is satisfied if any one of the following conditions holds:
\begin{itemize}
\item $\log f_{\mathbf{X}}(\mathbf{x})$ is Lipschitz over $\mathcal{X}$, and  $\mathcal{X}$ is $\mathbb{R}^n$ or has a measure-zero boundary.
\item $\mathbf{X}$ follows a multivariate Gaussian distribution with a full-rank covariance matrix.
\end{itemize}
\end{prop}

\medskip{}

\ifshortver
Moreover, the RSUQ retains the universal quantization property of \cite{ziv1985universal,zamir1992universal},
i.e., it is an almost-optimal lossy compressor for any distribution
of the input $\mathbf{X}$, for a general difference distortion measure $d:\mathbb{R}^{n}\to[0,\infty)$ (where the distortion between $\mathbf{X}$
and $\mathbf{Y}$ is $d(\mathbf{Y}-\mathbf{X})$). 
The following is a generalization of \cite{zamir1992universal} to RSUQ, proved using similar techniques as
in \cite{zamir1992universal}. The proof is given in  
\cite{ling2024rejectionsampled}.
\else
Moreover, the RSUQ retains the universal quantization property of \cite{ziv1985universal,zamir1992universal},
i.e., it is an almost-optimal lossy compressor for any distribution
of the input $\mathbf{X}$. Now, instead of the squared $2$-norm distortion,
we may consider a more general difference distortion measure. Let $d:\mathbb{R}^{n}\to[0,\infty)$
be a distortion function. The distortion between the input $\mathbf{X}$
and output $\mathbf{Y}$ is $d(\mathbf{Y}-\mathbf{X})$. 

The following result is a generalization of the universal quantization
result in \cite{zamir1992universal} to RSUQ, showing that the conditional
entropy of RSUQ is close to the rate-distortion function, regardless
of the input distribution. It is proved using similar techniques as
in \cite{zamir1992universal}. The proof is given in Appendix~\ref{subsec:pf_universal}.
\fi

\myskip

\begin{thm}\label{thm:universal}
Consider the rejection-sampled universal quantizer $Q_{\mathcal{A},\mathcal{P}}$
for $\mathcal{A}$ against $\mathcal{P}$. For a random input $\mathbf{X}$
following any distribution, independent of $S\sim P_{S}$, the conditional
entropy of the quantizer satisfies
\[
H(Q_{\mathcal{A},\mathcal{P}}(\mathbf{X},S)\,|\,S)\le R(D)-\log p+C(D)+\log e,
\]
where $p:=\mu(\mathcal{A})/\mu(\mathcal{P})$, $D:=\mathbb{E}[d(Q(\mathbf{X},S)-\mathbf{X})]$
is the expected distortion of the RSUQ, $R(D):=\inf_{P_{\hat{\mathbf{X}}|\mathbf{X}}:\,\mathbb{E}[d(\hat{\mathbf{X}}-\mathbf{X})]\le D}I(\mathbf{X};\hat{\mathbf{X}})$
is the rate-distortion function, 
\iflongver
and
\[
C(D):=\sup_{P_{\mathbf{T}}:\,\mathbb{E}[d(-\mathbf{T})]\le D}I(\mathbf{T};\mathbf{T}+\mathbf{Z}),
\]
where the supremum is over $P_\mathbf{T}$ (let $\mathbf{T} \sim P_\mathbf{T}$) satisfying $\mathbb{E}[d(-\mathbf{T})]\le D$, and $\mathbf{Z}\sim\mathrm{Unif}(\mathcal{A})$.
\else
$C:=\sup_{P_{\mathbf{T}}:\,\mathbb{E}[d(-\mathbf{T})]\le D}I(\mathbf{T};\mathbf{T}+\mathbf{Z})$
where $\mathbf{Z}\sim\mathrm{Unif}(\mathcal{A})$.
\fi
\end{thm}
\myskip

For squared error distortion $d(\mathbf{x}) = \Vert \mathbf{x}\Vert^2$, $C(D)$ is bounded in terms of the normalized second moment~\eqref{eq:nsm} of $\mathcal{A}$ (see \cite{zamir1992universal}):
\ifshortver
$ C(D) \le (n/2)\log (4 \pi e G_n(\mathcal{A}))$. 
\else
\begin{align*}
& C(D) \le \frac{n}{2}\log \big(4 \pi e G_n(\mathcal{A})\big). 
\end{align*}
\fi

\medskip{}

\section{Nonuniform Error Distribution} \label{sec:LRSUQ}

\iflongver
We have presented the RSUQ construction that guarantees a uniform
error distribution over a bounded set $\mathcal{A}$. 
In this section,
we generalize the construction to general (uniform or nonuniform)
continuous error distributions. 
This is based on the idea in \cite{wilson2000layered,hegazy2022randomized}
that any continuous distribution can be expressed as a mixture of
uniform distributions
(also see \cite{agustsson2020universally} for the Gaussian case).
\else
We generalize the construction to general (uniform or nonuniform)
continuous error distributions, using the idea in \cite{wilson2000layered,hegazy2022randomized}
that any continuous distribution can be expressed as a mixture of
uniform distributions
(also see \cite{agustsson2020universally} for the Gaussian case).
\fi 
Consider a 
probability density function $f:\mathbb{R}^{n}\to[0,\infty)$.
Write its superlevel set as
\iflongver
\[
L_{t}^{+}(f):=\{\mathbf{z}\in\mathbb{R}^{n}:\,f(\mathbf{z})\ge t\}.
\]
\else
$L_{t}^{+}(f):=\{\mathbf{z}\in\mathbb{R}^{n}:\,f(\mathbf{z})\ge t\}$.
\fi
Let $f_{T}(t):=\mu(L_{t}^{+}(f))$ for $t>0$, which is also a probability
density function. If we generate $T\sim f_{T}$, and then $\mathbf{Z}|\{T=t\}\sim\mathrm{Unif}(L_{t}^{+}(f))$,
then we have $\mathbf{Z}\sim f$.\footnote{\label{footnote:layered}The fundamental theorem of simulation \cite{robert2004monte} states
that if we generate $(\mathbf{Z},T)\sim\mathrm{Unif}(\{(\mathbf{z},t):\,0\le t\le f(\mathbf{z})\})$,
then $\mathbf{Z}\sim f$. Here $T\sim f_{T}$, and the conditional
distribution of $\mathbf{Z}$ given $T=t$ is $\mathrm{Unif}(L_{t}^{+}(f))$.
\iflongver
Hence, we can generate $T$ first, and then generate $\mathbf{Z}$ given $T$.
\fi
} 
Therefore, we can express the distribution $f$ as a mixture of uniform
distributions $\mathrm{Unif}(L_{t}^{+}(f))$. We now generalize RSUQ
to allow a nonuniform error distribution $f$. We first generate
$T\sim f_{T}$ randomly as a part of the random state, and then apply
RSUQ for $L_{T}^{+}(f)$. 
This is a generalization of the layered
randomized quantizer in \cite{hegazy2022randomized} 
\iflongver
(which focuses on scalar quantization, i.e., $n=1$).
\else
(which focuses on $n=1$).
\fi

\medskip{}

\begin{defn}
\label{def:rej_samp_quant_layer}Given a basic cell $\mathcal{P}$
of the lattice $\mathbf{G}\mathbb{Z}^{n}$, and a probability density
function $f:\mathbb{R}^{n}\to[0,\infty)$ where $L_{t}^{+}(f)$ is
always bounded for $t>0$, and $\beta:(0,\infty)\to[0,\infty)$ satisfying
$L_{t}^{+}(f)\subseteq\beta(t)\mathcal{P}$ for $t>0$, the \emph{layered
rejection-sampled universal quantizer} (LRSUQ) for $f$ against $\mathcal{P}$
is the randomized quantizer $(P_{S},Q_{f,\mathcal{P}})$, where $S=(T,(\mathbf{V}_{i})_{i\in\mathbb{N}^{+}})$,
$T\sim f_{T}$ where $f_{T}(t):=\mu(L_{t}^{+}(f))$, and $\mathbf{V}_{1},\mathbf{V}_{2},\ldots\stackrel{iid}{\sim}\mathrm{Unif}(\mathcal{P})$
\ifshortver
are i.i.d. dither signals. The quantization function
\else
is a sequence of i.i.d. dither signals. The quantization function is
\fi
\[
Q_{f,\mathcal{P}}(\mathbf{x},t,(\mathbf{v}_{i})_{i}):=\beta(t)\cdot\big(Q_{\mathcal{P}}(\mathbf{x}/\beta(t)-\mathbf{v}_{k})+\mathbf{v}_{k}\big),
\]
\iflongver
where
\fi
\begin{equation*}
k:=\min\big\{ i:\,\beta(t)\cdot\big(Q_{\mathcal{P}}(\mathbf{x}/\beta(t)-\mathbf{v}_{i})+\mathbf{v}_{i}\big)-\mathbf{x}\in L_{t}^{+}(f)\big\},
\end{equation*}
and $Q_{\mathcal{P}}$ is the lattice quantizer for basic cell
$\mathcal{P}$ (Section \ref{sec:lattice}).\footnote{Since $\beta(T)>0$ almost surely, division by zero will not occur
in $\mathbf{x}/\beta(t)$.}
\end{defn}

\myskip

\ifshortver
Properties of the LRSUQ are discussed in 
the preprint
\cite{ling2024rejectionsampled} due to space constraint. In particular, among randomized quantizers with error distribution $f$, the LRSUQ attains a normalized entropy within $\log e$ bits from the layered entropy lower bound \cite{hegazy2022randomized}. Hence, we can characterize the ``high SNR limit'' of one-shot channel simulation \cite{bennett2002entanglement,harsha2010communication,sfrl_trans,hegazy2022randomized} for additive noise channels within $\log e$ bits.
\else
The encoding and decoding operations of LRSUQ are given in  Algorithms~\ref{alg:LRSUQ_encode} and~\ref{alg:LRSUQ_decode}, respectively.

\begin{algorithm}[ht]
\textbf{$\;\;\;\;$Input:} vector  $\mathbf{x}\in\mathbb{R}^n$, full-rank generator matrix $\mathbf{G} \in \mathbb{R}^{n\times n}$, basic cell $\mathcal{P}$ of $\mathbf{G}\mathbb{Z}^n$, pdf $f$, function $\beta:(0,\infty)\to[0,\infty)$ satisfying
$L_{t}^{+}(f)\subseteq\beta(t)\mathcal{P}$ for $t>0$, random number generator (RNG) $\mathfrak{B}$

\textbf{$\;\;\;\;$Output:} description $(K,
\mathbf{M}) \in \mathbb{N}^{+} \times \mathbf{G}\mathbb{Z}^n$ 

\smallskip{}

\begin{algorithmic}[1]
\State Sample $T \sim f_T$ where $f_{T}(t):=\mu(L_{t}^{+}(f))$ using RNG $\mathfrak{B}$ 
\For{$i=1, 2, \ldots$} 
\State \label{step:LRS1} Sample 
$\mathbf{V}_{i} \sim\mathrm{Unif}(\mathcal{P})$ using RNG $\mathfrak{B}$
\State $\mathbf{M} \leftarrow Q_{\mathcal{P}}(\mathbf{x}/\beta(T)-\mathbf{V}_{i})\in\mathbf{G}\mathbb{Z}^{n}$ 
\If{$\beta(T)\cdot(\mathbf{M}+\mathbf{V}_i)-\mathbf{x}\in L_{T}^{+}(f)$} 
\State \Return $(i,\mathbf{M})$
\EndIf
\EndFor
\end{algorithmic}

\caption{\label{alg:LRSUQ_encode}$\textsc{LRSUQ\_Encode}(\mathbf{x},\mathbf{G},\mathcal{P},f,\beta(t),\mathfrak{B})$}
\end{algorithm}

\begin{algorithm}[ht]
\textbf{$\;\;\;\;$Input:} description $(K,\mathbf{M}) \in \mathbb{N}^{+}\times \mathbf{G}\mathbb{Z}^n$, full-rank generator matrix $\mathbf{G} \in \mathbb{R}^{n\times n}$, basic cell $\mathcal{P}$ of $\mathbf{G}\mathbb{Z}^n$, pdf $f$, function $\beta:(0,\infty)\to[0,\infty)$ satisfying
$L_{t}^{+}(f)\subseteq\beta(t)\mathcal{P}$ for $t>0$, RNG $\mathfrak{B}$

\textbf{$\;\;\;\;$Output:} $
\mathbf{Y} \in \mathbb{R}^{n} $ 

\smallskip{}

\begin{algorithmic}[1]
\State Sample $T \sim f_T$ where $f_{T}(t):=\mu(L_{t}^{+}(f))$ using RNG $\mathfrak{B}$ 
\State Sample $\mathbf{V}_{1},\ldots,\mathbf{V}_{K}\stackrel{iid}{\sim}\mathrm{Unif}(\mathcal{P})$ using RNG $\mathfrak{B}$ 
\State \Return  $ \beta(T)\cdot(\mathbf{M}+\mathbf{V}_{K})$ 
\end{algorithmic}

\caption{\label{alg:LRSUQ_decode}$\textsc{LRSUQ\_Decode}((K,\mathbf{M}),\mathbf{G},\mathcal{P},f,\beta(t),\mathfrak{B})$}
\end{algorithm}

First, we show that the LRSUQ indeed gives the desired error distribution.

\medskip
\begin{prop}
Consider the layered rejection-sampled universal quantizer $Q_{f,\mathcal{P}}$. 
For any random input $\mathbf{X}$, the quantization error $\mathbf{Z} := Q_{f,\mathcal{P}}(\mathbf{X},T,(\mathbf{V}_{i})_{i}) - \mathbf{X}$ follows the pdf $f$, independent of $\mathbf{X}$.
\end{prop}
\medskip
\begin{IEEEproof}
If $T=t$ is fixed and $(\mathbf{V}_{i})_{i\in \mathbb{N}^{+}}$ is still random, then the LRSUQ becomes the RSUQ in Definition~\ref{def:rej_samp_quant} with the basic cell $\beta(t)\mathcal{P}$.
We have $\mathbf{Z}\,|\,\{T=t\} \sim \mathrm{Unif}(L_{t}^{+}(f))$ independent of $\mathbf{X}$ by Proposition~\ref{prop:error_dist_unif}.
Hence, $\mathbf{Z} \sim f$ by the fundamental theorem of simulation \cite{robert2004monte} (footnote \ref{footnote:layered}).
\end{IEEEproof}
\medskip{}

The LRSUQ can also be regarded as a one-shot channel simulation scheme
\cite{bennett2002entanglement,harsha2010communication,sfrl_trans},
where the encoder observes $\mathbf{X}$, and sends a sequence of
bits to the decoder (which can share common randomness with the encoder)
to allow it to generate $\mathbf{Y}$ following a prescribed conditional
distribution $P_{\mathbf{Y}|\mathbf{X}}$ given $\mathbf{X}$. LRSUQ
is a one-shot channel simulation scheme for the additive noise channel
$\mathbf{Y}=\mathbf{X}+\mathbf{Z}$ where $\mathbf{Z}\sim f$.

\medskip{}

To study the normalized entropy of LRSUQ, we utilize the following quantity introduced in \cite{hegazy2022randomized}.

\myskip

\begin{defn}
[\cite{hegazy2022randomized}]\label{def:layered_ent}Given a probability
density function $f:\mathbb{R}^{n}\to[0,\infty)$, its \emph{layered
entropy} is
\[
h_{L}(f):=\int_{0}^{\infty}\mu(L_{t}^{+}(f))\log\mu(L_{t}^{+}(f)) \mathrm{d}t.
\]
\end{defn}
\medskip{}

The layered entropy has the following alternative characterization
(the one-dimensional case has been used as an intermediate step in
\cite{hegazy2022randomized}). The proof is given in Appendix \ref{subsec:pf_layered_alt}.
\medskip
\begin{thm}
\label{thm:layered_alt}We have
\[
h_{L}(f)=\sup h(\mathbf{Z}|T)
\]
where $\mathbf{Z}\sim f$, and the supremum is over conditional distributions
$P_{T|\mathbf{Z}}$ (let $T|\mathbf{Z}\sim P_{T|\mathbf{Z}}$) satisfying
that the conditional distribution $P_{\mathbf{Z}|T}(\cdot|t)$ is
uniform (i.e., $P_{\mathbf{Z}|T}(\cdot|t)$ is $\mathrm{Unif}(\mathcal{A}_{t})$
for some $\mathcal{A}_{t}\subseteq\mathbb{R}^{n}$ with $0<\mu(\mathcal{A}_{t})<\infty$)
for $P_{T}$-almost all $t$. Moreover, the supremum is attained when
$(\mathbf{Z},T)\sim\mathrm{Unif}(\{(\mathbf{z},t):\,0\le t\le f(\mathbf{z})\})$.
\end{thm}
\medskip{}

As a result of Definition~\ref{def:layered_ent} and Theorem~\ref{thm:layered_alt},\footnote{The upper bound $h_L(f) \le h(f)$ when $f$ is a one-dimensional unimodal probability density function was shown in~\cite{hegazy2022randomized}.}
\begin{align}
h_{\infty}(f) \le h_L(f) \le h(f), \label{eq:hl_bounds}
\end{align}
where $h_{\infty}(f) := - \log \mathrm{ess} \sup_{\mathbf{z}} f(\mathbf{z})$ is the differential R{\'{e}}nyi entropy of order $\infty$~\cite{renyi1961measures,ram2016renyi}.

It was shown in \cite{hegazy2022randomized} that, for a one-dimensional
error distribution $f:\mathbb{R}\to[0,\infty)$ that is unimodal (i.e.,
$L_{t}^{+}(f)$ is always connected), among randomized quantizers
that guarantees an error distribution $f$, the smallest possible
normalized entropy is given by $-h_{L}(f)$. The optimum is achieved
by the layered randomized quantizer \cite{hegazy2022randomized}.
In this paper, we will show that for a general number of dimensions
$n$, and a general continuous distribution $f$ over $\mathbb{R}^{n}$
(not necessarily unimodal), the smallest possible normalized entropy
for the error distribution $f$ is still given by $-h_{L}(f)$ within
a constant gap of $\log e$. The proof is divided into two parts:
the achievability part showing that LRSUQ has a normalized entropy
upper-bounded by $-h_{L}(f)+\log e$, and the converse part showing
that no randomized quantizer with error distribution $f$ can have
a normalized entropy smaller than $-h_{L}(f)$.
The proof is given in Appendix~\ref{subsec:pf_rsuq_ent_layer}.

\medskip{}

\begin{thm}
\label{thm:rsuq_ent_layer}Consider the layered rejection-sampled
universal quantizer $Q_{f,\mathcal{P}}$. For a random input $\mathbf{X}$
satisfying $\mathbf{X}\in\mathcal{X}$ almost surely for some $\mathcal{X}\subseteq\mathbb{R}^{n}$
with $\mu(\mathcal{X})<\infty$, independent of $S\sim P_{S}$, the
conditional entropy of the quantizer satisfies
\begin{align}
 & H(Q_{f,\mathcal{P}}(\mathbf{X},S)\,|\,S)\nonumber \\
 & \le\mathbb{E}_{T}\big[\log\mu(\mathcal{X}+3\eta\beta(T)B^{n})\big]-h_{L}(f)+\log e,\label{eq:HqXS-1}
\end{align}
where $T\sim f_{T}$ with $f_{T}(t):=\mu(L_{t}^{+}(f))$, $\eta:=\sup_{\mathbf{x}\in\mathcal{P}}\Vert\mathbf{x}\Vert$,
and $\beta(t)$ is given in Definition \ref{def:rej_samp_quant_layer}.
Hence, under the mild assumption that $\mathbb{E}[\log(1+\beta(T))]<\infty$,
the normalized entropy is upper-bounded by
\[
-h_{L}(f)+\log e.
\]
\end{thm}
\medskip{}

Note that we can also study the \emph{worst-case} normalized entropy
\begin{equation}
\underset{\tau \to\infty}{\lim\sup}\Big(\sup_{P_{\mathbf{X}}:\,\Vert\mathbf{X}\Vert\le \tau }H(Q(\mathbf{X},S)|S)-\log\mu(\tau B^n)\Big),\label{eq:worst_norment}
\end{equation}
which concerns the worst-case conditional entropy over all distributions
of $\mathbf{X}$ over $\tau B^n$ (instead of only $\mathbf{X}\sim\mathrm{Unif}(rB^{n})$
as in the original normalized entropy in Definition \ref{def:norment}).
Theorem \ref{thm:rsuq_ent_layer} establishes that the worst-case
normalized entropy of LRSUQ is also at most $-h_{L}(f)+\log e$ (under
the assumption $\mathbb{E}[\log(1+\beta(T))]<\infty$).

As a result, we can construct an LRSUQ for an arbitrary continuous
error distribution $f$ with normalized entropy at most $-h_{L}(f)+\log e$,
under a mild assumption on $f$.\footnote{The authors are unaware of any naturally occurring distribution $f$
that violates condition \eqref{eq:cor_f_cond}.}
The proof is given in Appendix~\ref{subsec:pf_rsuq_exist}.
\medskip{}

\begin{cor}
\label{cor:rsuq_exist}If a probability density function $f$ over
$\mathbb{R}^{n}$ satisfies 
\begin{equation}
\int_{0}^{\infty}\Big(\sup_{\mathbf{z}:\,\Vert\mathbf{z}\Vert\ge x}f(\mathbf{z})\Big)x^{n-1}\log(1+x)\mathrm{d}x<\infty,\label{eq:cor_f_cond}
\end{equation}
then there exists an LRSUQ $Q_{f,\mathcal{P}}$ for $f$ with normalized
entropy upper-bounded by $-h_{L}(f)+\log e$.
\end{cor}

\mysmallskip

We then show the converse part, which generalizes the scalar randomized
quantization converse result in \cite{hegazy2022randomized} to vector
quantization. The proof is given in Appendix \ref{subsec:pf_nonunif_converse}.

\mysmallskip

\begin{thm}
\label{thm:nonunif_converse}If a randomized quantizer satisfies that
the error $Q(\mathbf{X},S)-\mathbf{X}$ has a probability density
function $f$ when $\mathbf{X}\sim\mathrm{Unif}(\mathcal{X})$ for
a certain set $\mathcal{X}\subseteq\mathbb{R}^{n}$, $0<\mu(\mathcal{X})<\infty$,
then we have the following bound on the conditional entropy
\[
H(Q(\mathbf{X},S)|S)\ge\log\mu(\mathcal{X})-h_{L}(f).
\]
As a result, if the error $Q(\mathbf{x},S)-\mathbf{x}\sim f$ for
every $\mathbf{x}\in\mathbb{R}^{n}$, then the normalized entropy
of the randomized quantizer is lower-bounded by $-h_{L}(f)$.
\end{thm}
\medskip{}

Corollary \ref{cor:rsuq_exist} and Theorem \ref{thm:nonunif_converse}
establish that, letting $\overline{H}^{*}(f)$ be the smallest\footnote{Technically, $\overline{H}^{*}(f)$ is the infimum of the set of normalized
entropies of randomized quantizers with error distribution $f$.} normalized entropy among randomized quantizers with error distribution
$f$ (i.e., $Q(\mathbf{x},S)-\mathbf{x}\sim f$ for every $\mathbf{x}$),
under the assumption \eqref{eq:cor_f_cond}, we have
\begin{equation}
-h_{L}(f)\le\overline{H}^{*}(f)\le-h_{L}(f)+\log e.\label{eq:hl_approx}
\end{equation}
Therefore, $h_{L}(f)$ approximately characterizes the ``high SNR limit'' of one-shot additive noise channel simulation, that is,
the communication cost for
simulating 
the channel $\mathbf{X}\to\mathbf{X}+\mathbf{Z}$ (where
$\mathbf{Z}\sim f$, $\mathbf{X}\sim\mathrm{Unif}(rB^{n})$)\footnote{We can also study the worst-case cost for simulating the channel $\mathbf{X}\to\mathbf{X}+\mathbf{Z}$,
where $\mathbf{X}$ follows any distribution with $\Vert\mathbf{X}\Vert\le r$.
The construction in Theorem \ref{thm:rsuq_ent_layer} establishes
that the cost is still approximately $\log\mu(rB^{n})-h_{L}(f)$.} is approximately $\log\mu(rB^{n})-h_{L}(f)$ under the high SNR limit $r \to \infty$.
We will see that each of the other two entropies in \eqref{eq:hl_bounds} corresponds to a high SNR limit of channel simulation under certain assumptions as well.

The differential entropy $h(f)$ characterizes the high SNR limit of the asymptotic capacity of the additive noise channel
$\mathbf{X}\to\mathbf{X}+\mathbf{Z}$ (under the input constraint $\Vert\mathbf{X}\Vert\le r$). As $r\to\infty$, the capacity is approximately $\log\mu(rB^{n})-h(f)$. By the reverse Shannon theorem~\cite{bennett2002entanglement}, this is the rate needed for the asymptotic simulation of this channel, i.e., we simulate the $N$-fold memoryless channel $(\mathbf{X}_i)_{i=1}^N \to (\mathbf{X}_i+\mathbf{Z}_i)_{i=1}^N $ and let the blocklength $N \to \infty$. Therefore, if we keep the blocklength $N=1$ and only let $r\to \infty$ as in LRSUQ (Definition~\ref{def:rej_samp_quant_layer}), then we require $\approx \log\mu(rB^{n})-h_{L}(f)$ bits; whereas if we also let $N\to \infty$, then we require a smaller rate $\log\mu(rB^{n})-h(f)$. However, the asymptotic assumption $N\to \infty$ is often not realistic, and the one-shot nature of LRSUQ makes it more applicable to practical scenarios.

The order-$\infty$ differential R{\'{e}}nyi entropy $h_{\infty}(f) = - \log \mathrm{ess} \sup_{\mathbf{z}} f(\mathbf{z})$ characterizes the high SNR limit of one-shot channel simulation if we use a straightforward rejection sampling scheme instead of the LRSUQ. Consider the RSUQ in Definition~\ref{def:rej_samp_quant} where, instead of performing rejection sampling for the distribution $\mathrm{Unif}(\mathcal{A})$ against $\mathrm{Unif}(\mathcal{P})$, we now perform rejection sampling for the probability density function $f$ (supported over $\mathcal{P}$) against $\mathrm{Unif}(\mathcal{P})$. The acceptance probability is then $p = (\mu(\mathcal{P}) \mathrm{ess} \sup_{\mathbf{z}} f(\mathbf{z}))^{-1}$. By the same arguments as in Theorem~\ref{thm:rsuq_ent}, the normalized entropy of this scheme is upper-bounded by $-h_{\infty}(f) + \log e$. Since $h_L(f) \ge h_{\infty}(f)$, the bound on the normalized entropy of this simple scheme is weaker than that of LRSUQ.

\medskip{}

\fi

\begin{rem} \label{remark:time} 
To realize the channel simulation in practice, the computation complexity, i.e., the number of samples to be generated, is a major factor. It is shown in \cite{agustsson2020universally} that given any target and proposal distributions, under the assumption $\mathrm{RP} \neq \mathrm{NP}$, there do not exist a universal algorithm with a computational complexity that is polynomial in terms of KL divergence between the distributions. 
Moreover, Goc and Flamich  \cite{goc2024channel} showed that the expected running time of the class of CRS algorithms is at least exponential in terms of the R\'{e}nyi-$\infty$ divergence between the proposal and target distributions.
Thus, the computational cost of a scheme for channel simulation is distribution-dependent.
For instance, \cite{agustsson2020universally} 
discussed schemes for simulating channels with i.i.d. additive uniform noises or Gaussian noises with a time complexity linear in the dimension.
In contrast, the expected number of steps of RSUQ to simulate an additive uniform channel over an arbitrary set is inversely proportional to the packing density $\delta(\mathbf{G})$ that grows exponentially.
The computational complexity of LRSUQ applied to a specified target distribution, such as a multivariate Gaussian distribution, is left for future study. 
\end{rem}

\subsection{Excess information of LRSUQ}

Recall that for lossy compression via RSUQ, the gap between the conditional entropy of RSUQ and the theoretical limit was measured by the redundancy \eqref{eq:red_maxerror}, \eqref{eq:red_mse_shannon} and \eqref{eq:red_mse_zador}. For channel simulation via LRSUQ, the gap between the conditional entropy and the theoretical limit can be measured by the excess information~\cite{sfrl_trans}, which is the difference between the conditional entropy and the mutual information $I(\mathbf{X};\mathbf{Y})$ between the input and the output, which lower-bounds the conditional entropy of any channel simulation scheme that simulates the channel from $\mathbf{X}$ to $\mathbf{Y}$ exactly. In the context of this paper, the \emph{normalized excess information} (in the high SNR limit) of a randomized quantizer $Q$ for the simulation of the channel $P_{\mathbf{Y}|\mathbf{X}}$ is given by
\begin{align}
\Phi(Q) := \frac{1}{n}\underset{\tau\to\infty}{\lim\sup}\big(H(Q(\mathbf{X}_{\tau},S)|S)-I(\mathbf{X}_{\tau};\mathbf{Y}_{\tau})\big), \label{eq:norm_excess}
\end{align}
where $\mathbf{X}_{\tau}\sim\mathrm{Unif}(\tau B^n)$ is independent of
$S\sim P_{S}$, and $\mathbf{Y}_{\tau}$ follows $P_{\mathbf{Y}|\mathbf{X}}$ given $\mathbf{X}_{\tau}$. For randomized quantizers that exactly simulates $P_{\mathbf{Y}|\mathbf{X}}$ such as LRSUQ, we can take $\mathbf{Y}_{\tau} = Q(\mathbf{X}_{\tau},S)$ to be the output of the randomized quantizer.
The normalized excess information is normalized by dividing the dimension $n$, as in \cite{KobusRotated2024}. 

We then characterize the normalized excess information $\Phi(Q)$ in terms of the normalized entropy $\overline{H}(Q)$ for additive noise channels. 
The proof is given in Appendix~\ref{subsec:pf_lb_normexcees}.

\medskip

\begin{prop}\label{prop:norm_excess_additive}
If $P_{\mathbf{Y}|\mathbf{X}}$ is an additive noise channel $\mathbf{Y} = \mathbf{X} + \mathbf{Z}$ where $\mathbf{Z} \sim f$ with finite second moment, then
\begin{align}
\Phi(Q) = \frac{1}{n}\big(\overline{H}(Q) + h(f) \big). \label{eq:norm_excess_additive}
\end{align}
As a result, the smallest possible $\Phi(Q)$ among randomized quantizers for the exact simulation of the additive noise channel with error $\mathbf{Z}\sim f$ is
\[
\inf_{Q} \Phi(Q) = \frac{1}{n}\big(\overline{H}^{*}(f) + h(f) \big),
\]
where the infimum is over randomized quantizers with error distribution $f$. Using \eqref{eq:hl_approx}, we have\footnote{The normalized excess information $\Phi(Q)$ is lower-bounded by $\frac{1}{n}\big(h(f) -h_{L}(f) \big)$ which is generally positive. This suggests that instead of using $\Phi(Q)$, we may use a tighter definition of redundancy $\frac{1}{n}\big(\overline{H}(Q) + h_L(\mathbf{Z}) \big)$ which must also be nonnegative, where we use the layered entropy $h_L(\mathbf{Z})$ instead of the differential entropy $h(\mathbf{Z})$ in \eqref{eq:norm_excess_additive}.}
\begin{equation}
\frac{1}{n}\big(h(f) -h_{L}(f) \big) \le \inf_{Q} \Phi(Q) \le \frac{1}{n}\big(h(f) -h_{L}(f) + \log e \big). \label{eq:norm_excess_bounds}
\end{equation}
\end{prop}

\medskip

\begin{table*}[ht]
\caption{
Comparison of \( h_{L}(\mathbf{Z}) \) for \( \mathbf{Z} \sim \mathcal{N}(\mathbf{0}, \mathbf{I}_{n\times n}) \) (2nd column), lower bounds on \( \Phi(Q) \) for any randomized quantizer simulating \\ the Gaussian channel (3rd column), \( \Phi(Q) \) of LRSUQ (4th column), \( \Phi(Q) \) of LSPQ (5th column), and excess information of RDQ \\ (6th column), across dimensions \( n \).}
\label{tab:excess_LRSUQ}
\centering
\sisetup{table-format=2.5, table-number-alignment=center}
\begin{tabular}{@{} c S[table-format=2.5] S S S S[table-format=1.5(3)] @{}}
\toprule
\textbf{Dimension} & 
\multicolumn{1}{c}{\textbf{\( h_{L}(\mathbf{Z}), \mathbf{Z} \sim \mathcal{N}(\mathbf{0}, \mathbf{I}_{n\times n}) \)}} & 
{\textbf{Lower Bound on $\Phi(Q)$}} & 
{\textbf{$\Phi(Q)$ of  LRSUQ}} & 
{\textbf{$\Phi(Q)$ of LSPQ}} & 
{\textbf{Excess Information of RDQ}} \\
\( n \) & 
{(bits)} & 
{ (bits/dim)} & 
{(bits/dim)} & 
{(bits/dim)} & 
{(bits/dim)} \\
\midrule
1    & 1.52632 & 0.52077 & 0.52077\textsuperscript{*} & 6.13777 & 0.25461 \\ 
2    & 3.26144 & 0.41637 & 1.13772          & 4.03337 & 0.22686 \\
3    & 5.08819 & 0.35103 & 0.83193          & 3.30136 & 0.21376 \\
4    & 6.96559 & 0.30570 & 0.66637          & 2.92270 & 0.19387 \\
5    & 8.87490 & 0.27212 & 0.56065          & 2.68912 & 0.18613 \\
6    & 10.80611 & 0.24608 & 0.48653         & 2.52974 & 0.17230 \\
7    & 12.75325 & 0.22520 & 0.43130         & 2.41363 & 0.16139 \\
8    & 14.71250 & 0.20803 & 0.38837         & 2.32503 & 0.14597 \\
24   & 46.71338 & 0.10070 & 0.16082          & 1.88437 & 0.08389 $\pm$ 0.00081 \\
\bottomrule
\end{tabular}
\vspace{0.2cm}
\footnotesize{\\ \textsuperscript{*} No \( \log e \) term.}
\end{table*}

\begin{figure*}[htbp]
\begin{centering}
\includegraphics[scale=0.55]{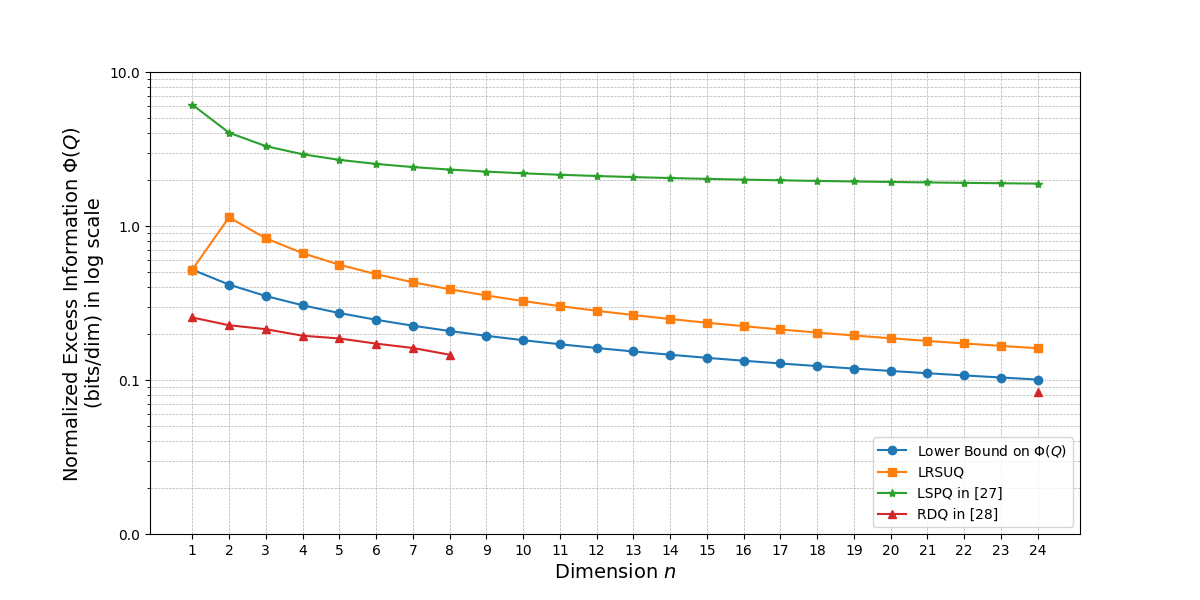} \hspace{16pt} 
\par\end{centering}
\caption{\label{fig:ExcessInfo_compare} Comparison of the lower bound on $\Phi(Q)$ for any randomized quantizer  simulating the Gaussian channel (blue curve), $\Phi(Q)$ of LRSUQ  (orange curve), $\Phi(Q)$ of LSPQ introduced in \cite{ling2023vector} (green curve), and   excess information of RDQ proposed in \cite{KobusRotated2024} (red curve), across various dimensions.}
\vspace{-3pt}
\end{figure*}

We now study the additive white Gaussian noise (AWGN) channel, where $\mathbf{Z} \sim \mathcal{N}(\mathbf{0},\mathbf{I}_{n \times n})$. In this case, the layered entropy is given by
\begin{align}
h_{L}(f_{\mathbf{Z}}) &= \int_0^{\infty} \Big(\frac{v}{2}\Big)^{n/2} e^{-v/2} \nonumber\\ 
\,&\quad\cdot  \frac{(n/2) \log(\pi v) - \log \Gamma(n/2+1)}{2 \Gamma(n/2+1)}  \mathrm{d}v. \label{eq:layered_ent_gaussian}
\end{align}
Refer to Appendix~\ref{subsec:layeredEnt_Gaussian} for the proof.
In Table~\ref{tab:excess_LRSUQ} and Figure~\ref{fig:ExcessInfo_compare}, we numerically evaluate \eqref{eq:layered_ent_gaussian} for various dimension $n$, and compare the following quantities:
\begin{itemize}
\item The lower bound of the normalized excess information $\Phi(Q)$ that must be satisfied by every randomized quantizer which simulates the AWGN exactly (the lower bound in \eqref{eq:norm_excess_bounds}).
\item The upper bound of $\Phi(Q)$ of the LRSUQ for the AWGN channel (the upper bound in \eqref{eq:norm_excess_bounds}).
\item The normalized excess information of the layered shift-periodic quantizer (LSPQ) in \cite{ling2023vector}, which is a different method from LRSUQ for the exact simulation of the AWGN channel.\footnote{The layered shift-periodic quantizer \cite[Section V]{ling2023vector} can be used to simulate the Gaussian channel $\mathbf{X} \to \mathbf{X}+\mathbf{Z}$, where $\mathbf{Z} \sim \mathcal{N}(\mathbf{0}, \mathbf{I}_{n \times n})$, using the layering construction \cite{hegazy2022randomized,wilson2000layered} with dithering. By applying the upper bound from \cite[Corollary 13]{ling2023vector}, we compute an upper bound on the normalized entropy of the layered shift-periodic quantizer, yielding $1.617n + 4 - h_{L}(f_{\mathbf{Z}})$ bits.}
\item The excess information of the rotated dithered quantizer (RDQ) in \cite{KobusRotated2024}, which is a method for the approximate simulation of the AWGN channel.
\end{itemize}
Note that when $n=1$, the LRSUQ coincides with the layered randomized quantizer in \cite{agustsson2020universally,hegazy2022randomized}. 
Consequently, the rejection sampling step becomes unnecessary, and the $\log e$ term is omitted.

We observe that the excess information of LRSUQ approaches the lower bound as $n \to \infty$. The rotated dithered quantizer achieves a lower $\Phi(Q)$ than LRSUQ. We note that the $\Phi(Q)$ of the rotated dithered quantizer (which is an approximate simulation scheme) is also lower than the lower bound (that must be satisfied for exact simulation), which means that part of the lower communication cost of rotated dithered quantizer comes from the fact that it only approximately simulates the AWGN channel. In comparison, LRSUQ simulates the AWGN exactly,
making LRSUQ suitable for scenarios where attaining the exact distribution is crucial, e.g., the Gaussian mechanism 
in differential privacy \cite{hasirciouglu2024communication,hegazy2024compression}.

\medskip

\iflongver
\section{Conclusion and Discussions}
In this paper, we studied the rejection-sampled universal quantizer (RSUQ), which has smaller maximum and mean-squared errors compared to currently known lattice quantizers \cite{sikiric2008generalization,agrell2023best} with the same entropy for moderately large dimensions in the high-resolution limit. 
Moreover, RSUQ possesses the desirable property, similar to the universal quantizer \cite{ziv1985universal,zamir1992universal}, that the quantization error is always uniform over a prescribed set (e.g., a ball) 
and independent of the input distribution.
Furthermore, given a prescribed nonuniform noise distribution under a regularity condition, RSUQ can be applied to simulate the additive channel with the noise distribution by utilizing the layered construction techniques described in \cite{wilson2000layered,hegazy2022randomized}, where the entropy is shown to be within a constant $\log e$ away from the optimum in the high SNR limit.

For potential future directions, it may be of interest to investigate the performance of RSUQ in 
machine learning \cite{havasi2019minimal,agustsson2020universally,flamich2020compressing} and differential privacy \cite{lang2022joint,shahmiri2023communication,hasirciouglu2024communication,hegazy2023compression,yan2023layered} using some lattices with fast decoding and quantization algorithms. 
Also, it may be of interest to study whether the efficiency of LRSUQ can be improved for simulating specific additive noise channels, such as the additive multivariate Gaussian noise channel.
\fi

\medskip

\section{Acknowledgements}

The authors would like to thank Iosif Pinelis who provides great insight to the proof in Appendix~\ref{subsec:pf_Nor_mix_chi2}. 
The authors would like to thank the anonymous reviewers of the conference version~\cite{Ling2024ISITRSUQ} and the associate editor and the anonymous reviewers of the full version for their valuable comments. The associate editor's advice to compare against Zador's lower bound, and the associate editor's and reviewers' suggestions to consider the recent development in lattice quantizers \cite{Kudryashov2010LowComplexLattice,Lyu2022BetterLatticeQuan,Agrell2025OptLatticeQuan,ordentlich2025voronoi}, have greatly improved the manuscript.

\medskip{}

\ifshortver
\else

\appendix

\subsection{Proof of Proposition \ref{prop:lb}\label{subsec:pf_lb}}

Consider a random $\mathbf{X}$, and let $\mathbf{Y}=Q(\mathbf{X},S)$.
Since $\mathbf{X}$ is independent of $S$, we have
\begin{align*}
H(\mathbf{Y}|S) & \ge I(\mathbf{X};\mathbf{Y}|S)\\
 & =I(\mathbf{X};\mathbf{Y},S)\\
 & \ge I(\mathbf{X};\mathbf{Y})\\
 & =h(\mathbf{X})-h(\mathbf{X}|\mathbf{Y})\\
 & \ge  h(\mathbf{X})-h(\mathbf{X}-\mathbf{Y}) \\
 &=h(\mathbf{X})-h(\mathbf{Y}-\mathbf{X}).
\end{align*}
If the
maximum error is at most $r$ (i.e., $\Vert\mathbf{Y}-\mathbf{X}\Vert\le r$),
then
\begin{align*}
H(\mathbf{Y}|S) & \ge h(\mathbf{X})-h(\mathbf{X}-\mathbf{Y})\\
 & \ge h(\mathbf{X})-\log\mu(rB^{n})\\
 & =h(\mathbf{X})-n\log r-\log\kappa_{n}.
\end{align*}
Substituting $\mathbf{X}\sim\mathrm{Unif}(\tau B^n)$ gives the lower
bound of the normalized entropy for fixed maximum error. 
If the mean squared error is at most $D$, since the Gaussian distribution
maximizes the differential entropy for a fixed mean squared norm,
\begin{align*}
H(\mathbf{Y}|S) & \ge h(\mathbf{X})-h(\mathbf{X}-\mathbf{Y})\\
 & \ge h(\mathbf{X})-\frac{n}{2}\log\frac{2\pi eD}{n}.
\end{align*}
Substituting $\mathbf{X}\sim\mathrm{Unif}(\tau B^n)$ and taking $\tau \to \infty$ gives Shannon's lower
bound of the normalized entropy for a given bound on the MSE. For Zador's lower bound,
consider $Q(\mathbb{R}^{n},s):=\{Q(\mathbf{x},s):\,\mathbf{x}\in\mathbb{R}^{n}\}$ and $Q^{-1}(\mathbf{y},s):=\{\mathbf{x}\in\mathbb{R}^{n}:\,Q(\mathbf{x},s)=\mathbf{y}\}$. Let $p_{\mathbf{Y}|S}(\mathbf{y}|s):=\mathbb{P}(\mathbf{Y}=\mathbf{y}|S=s)$. For $\mathbf{X}\sim\mathrm{Unif}(\tau B^{n})$, we have
\begin{align*}
 & \mathbb{E}\left[\Vert Q(\mathbf{X},S)-\mathbf{X}\Vert^{2}\right]\\
 & =\mathbb{E}\bigg[\sum_{\mathbf{y}\in Q(\mathbb{R}^{n},S)}\int_{Q^{-1}(\mathbf{y},s)\cap\tau B^{n}}\frac{\Vert\mathbf{y}-\mathbf{x}\Vert^{2}}{\tau^{n}\kappa_{n}}\mathrm{d}\mathbf{x}\bigg]\\
 & \stackrel{(a)}{\ge}\mathbb{E}\bigg[\sum_{\mathbf{y}\in Q(\mathbb{R}^{n},S)}\int_{(\mu(Q^{-1}(\mathbf{y},s)\cap\tau B^{n})/\kappa_{n})^{1/n}B^{n}}\frac{\Vert\mathbf{z}\Vert^{2}}{\tau^{n}\kappa_{n}}\mathrm{d}\mathbf{z}\bigg]\\
 & =\mathbb{E}\bigg[\sum_{\mathbf{y}\in Q(\mathbb{R}^{n},S)}\frac{n}{(n+2)\tau^{n}}\bigg(\frac{\mu(Q^{-1}(\mathbf{y},s)\cap\tau B^{n})}{\kappa_{n}}\bigg)^{1+2/n}\bigg]\\
 & =\frac{n\tau^{2}}{n+2}\mathbb{E}\bigg[\sum_{\mathbf{y}\in Q(\mathbb{R}^{n},S)}p_{\mathbf{Y}|S}(\mathbf{y}|S)^{1+2/n}\bigg]\\
 & =\frac{n\tau^{2}}{n+2}\mathbb{E}\bigg[\exp_{2}\bigg(\log\sum_{\mathbf{y}\in Q(\mathbb{R}^{n},S)}p_{\mathbf{Y}|S}(\mathbf{y}|S)^{1+2/n}\bigg)\bigg]\\
 & \stackrel{(b)}{\ge}\frac{n\tau^{2}}{n+2}\mathbb{E}\bigg[\exp_{2}\bigg(-\frac{2}{n}H(p_{\mathbf{Y}|S}(\cdot|S))\bigg)\bigg]\\
 & \stackrel{(c)}{\ge}\frac{n\tau^{2}}{n+2}\exp_{2}\bigg(-\frac{2}{n}H(\mathbf{Y}|S)\bigg),
\end{align*}
where (a) is because the minimum of $\int_{A}\Vert\mathbf{z}\Vert^{2}\mathrm{d}\mathbf{z}$ over measurable sets $A\subseteq\mathbb{R}^{n}$ with fixed $\mu(A)=t$ is attained when $A=(t/\kappa_{n})^{1/n}B^{n}$ \cite{conway2013sphere}, (b) is due to the inequality between Shannon entropy and R\'{e}nyi entropy, and (c) is by Jensen's inequality. Therefore, if $\mathbb{E}\left[\Vert Q(\mathbf{X},S)-\mathbf{X}\Vert^{2}\right]\le D'$, then
\begin{align*}
H(\mathbf{Y}|S)-\log\mu(\tau B^{n}) & \ge\frac{n}{2}\log\frac{n\tau^{2}}{(n+2)D'}-\log(\tau^{n}\kappa_{n})\\
 & =-\frac{n}{2}\log\frac{(n+2)D'}{n}-\log\kappa_{n}.
\end{align*}
The result follows from taking $D'\to D$.

\subsection{Proof of Proposition \ref{prop:error_dist_unif}\label{subsec:pf_error_dist_unif}}

In the definition of RSUQ, the dither signals are uniformly distributed over the basic cell $\mathcal{P}$, i.e., $\mathbf{V}_i \sim \mathrm{Unif}(\mathcal{P})$ for every $i \in \mathbb{N}^+$.
Fix any $\mathbf{x} \in \mathbb{R}^n$. 
Let $\mathbf{Y} := Q_{\mathcal{A},\mathcal{P}}(\mathbf{x}, (\mathbf{V}_i)_i) = Q_{\mathcal{P}}(\mathbf{x} - \mathbf{V}_K) + \mathbf{V}_K$ as in \eqref{eq:QAP_def}.
Let $\mathbf{Z}_i := Q_{\mathcal{P}}(\mathbf{x} - \mathbf{V}_i) + \mathbf{V}_i-\mathbf{x}$. By \cite[Lemma 4.1.1 and Theorem 4.1.1]{zamir2014}, we have $\mathbf{Z}_i \sim \mathrm{Unif}(\mathcal{P})$ i.i.d. for $i\in \mathbb{N}^+$. From \eqref{eq:kstar}, $K = \min\{i: \mathbf{Z}_i \in \mathcal{A}\}$, and hence $\mathbf{Z}_K = \mathbf{Y} - \mathbf{x} \sim \mathrm{Unif}(\mathcal{A})$ by the standard rejection sampling argument.

\subsection{Proof of Theorem \ref{thm:rsuq_ent}\label{subsec:pf_rsuq_ent}}

Let $\mathbf{m}:=Q_{\mathcal{P}}(\mathbf{x}-\mathbf{v}_{k})\in\mathbf{G}\mathbb{Z}^{n}$.
By \eqref{eq:kstar}, $\mathbf{m}+\mathbf{v}_{k}-\mathbf{x}\in\mathcal{A}$,
and hence $\mathbf{m}\in\mathcal{A}-\mathbf{v}_{k}+\mathbf{x}\subseteq\mathcal{A}-\mathcal{P}+\mathcal{X}$.
Let $\mathcal{M} := (\mathcal{X}+\mathcal{A}-\mathcal{P}) \cap \mathbf{G}\mathbb{Z}^{n}$.
We know $\mathbf{m} \in \mathcal{M}$.
We have $\mathcal{P}+\mathcal{M}\subseteq\mathcal{X}+\mathcal{A}+\mathcal{P}-\mathcal{P}$,
and $\mu(\mathcal{P}+\mathcal{M})=|\mathcal{M}|\mu(\mathcal{P})$. 

Consider a random $\mathbf{X}\in\mathcal{X}$. Let $K$ be defined
according to \eqref{eq:kstar}, and $\mathbf{M}:=Q_{\mathcal{P}}(\mathbf{X}-\mathbf{V}_{K})$.
Note that the output $Q_{\mathcal{P}}(\mathbf{X}-\mathbf{V}_{K})+\mathbf{V}_{K}$
is a function of $((\mathbf{V}_{i})_{i},K,\mathbf{M})$. Therefore
$H(Q_{\mathcal{A},\mathcal{P}}(\mathbf{X},(\mathbf{V}_{i})_{i})\,|\,(\mathbf{V}_{i})_{i})\le H(K)+H(\mathbf{M})$.
We have
\begin{align*}
H(\mathbf{M}) & =H(Q_{\mathcal{P}}(\mathbf{X}-\mathbf{V}_{K}))\\
 & \le\log|\mathcal{M}|\\
 & \le\log\mu(\mathcal{X}+\mathcal{A}+\mathcal{P}-\mathcal{P})-\log\mu(\mathcal{P}).
\end{align*}
For $K$, note that $K\sim\mathrm{Geom}(p)$, where $p:=\mu(\mathcal{A})/\mu(\mathcal{P})$
is the acceptance probability of the rejection sampling (see the proof of Proposition~\ref{prop:error_dist_unif}). By the entropy
of geometric random variable,
\begin{align*}
H(K) & =-\log p-\frac{1-p}{p}\log(1-p)\\
 & =-\log\mu(\mathcal{A})-\frac{1-p}{p}\log(1-p)+\log\mu(\mathcal{P})\\
 & \le-\log\mu(\mathcal{A})+\log e+\log\mu(\mathcal{P})
\end{align*}
since $-\frac{1-p}{p}\log(1-p)$ is decreasing for $p\in(0,1)$. The
result \eqref{eq:HqXS} follows from $H(Q_{\mathcal{A},\mathcal{P}}(\mathbf{X},(\mathbf{V}_{i})_{i})\,|\,(\mathbf{V}_{i})_{i})\le H(K)+H(\mathbf{M})$.
For the bound on the normalized entropy, assume $\sup_{\mathbf{x}\in\mathcal{P}}\Vert\mathbf{x}\Vert\le\eta$
(recall that a basic cell must be bounded) and substitute $\mathcal{X}=\tau B^n$,
we have $\mu(\mathcal{X}+\mathcal{A}+\mathcal{P}-\mathcal{P})\le\mu((\tau +3\eta)B^{n})=(1+o(1))\mu(\tau B^n)$
as $\tau \to\infty$, completing the proof.

\subsection{Proof of Corollary \ref{cor:rsuq_ball}\label{subsec:pf_rsuq_ball}}

We construct an RSUQ for the ball $rB^{n}$ against the Voronoi cell
$\mathcal{V}$ of some lattice $\gamma\mathbf{G}\mathbb{Z}^{n}$ (where
$\gamma>0$ is some scaling factor). To ensure $rB^{n}\subseteq\mathcal{V}$,
we need the radius of the ball to be not greater than the packing
radius of the lattice, i.e., $r\le\underline{\lambda}(\gamma\mathbf{G})=\gamma\underline{\lambda}(\mathbf{G})$.
Hence we can take $\gamma=r/\underline{\lambda}(\mathbf{G})$. This
RSUQ guarantees an error distribution $\mathbf{Y}-\mathbf{x}\sim\mathrm{Unif}(rB^{n})$,
and hence has a maximum error $r$.

Invoking Theorem \ref{thm:rsuq_ent}, the normalized entropy is upper-bounded
by
\[
-n\log r-\log\kappa_{n}-\frac{1-\delta(\mathbf{G})}{\delta(\mathbf{G})}\log(1-\delta(\mathbf{G})).
\]
This is because $\mu(rB^{n})=r^{n}\kappa_{n}$ and 
\[
\frac{\mu(rB^{n})}{\mu(\mathcal{V})}=\frac{r^{n}\kappa_{n}}{|\det((r/\underline{\lambda}(\mathbf{G}))\mathbf{G})|}=\delta(\mathbf{G}).
\]

Note that the MSE $\mathbb{E}[\Vert Q_{\mathcal{A},\mathcal{P}}(\mathbf{X},S)-\mathbf{X}\Vert^{2}]$
(for any distribution of $\mathbf{X}$) of the RSUQ for the ball $rB^{n}$
is
\begin{align}
 & \frac{1}{\mu(rB^{n})}\int_{rB^{n}}\Vert\mathbf{x}\Vert^{2}\mathrm{d}\mathbf{x}\nonumber \\
 & =r^{2}\frac{1}{\kappa_{n}}\int_{B^{n}}\Vert\mathbf{x}\Vert^{2}\mathrm{d}\mathbf{x}\nonumber \\
 & =r^{2}\frac{1}{\kappa_{n}}\int_{S^{n-1}}\int_{0}^{1}t^{n-1}\Vert t\mathbf{x}\Vert^{2}\mathrm{d}t\mathrm{d}\mathbf{x}\nonumber \\
 & =r^{2}\frac{1}{\kappa_{n}}\int_{S^{n-1}}\int_{0}^{1}t^{n+1}\mathrm{d}t\mathrm{d}\mathbf{x}\nonumber \\
 & =\frac{nr^{2}}{n+2},\label{eq:rsuq_mse}
\end{align}
where $S^{n-1}:=\{\mathbf{x}\in\mathbb{R}^{n}:\,\Vert\mathbf{x}\Vert=1\}$
is the $(n-1)$-dimensional sphere. An upper bound of the normalized
entropy for unit MSE can be obtained by substituting $r=\sqrt{1+2/n}$
(so that the MSE is $1$) into \eqref{eq:rsuq_ent_ub}.

\subsection{Bounding the gap between \eqref{eq:red_mse_shannon} and \eqref{eq:red_mse_zador}, and the Shannon redundancy of RSUQ\label{subsec:pf_mse_gap}}

The gap between \eqref{eq:red_mse_shannon} and \eqref{eq:red_mse_zador} is
\begin{align*}
 & \frac{1}{2}\log\frac{2\pi e}{n+2}-\frac{1}{n}\log\kappa_{n}\\
 & =\frac{1}{2}\log\frac{2e}{n+2}+\frac{1}{n}\log\Gamma(n/2+1)\\
 & \stackrel{(a)}{=}\frac{1}{2}\log\frac{2e}{n+2}+\frac{1}{2}\log\frac{n}{2e}+\frac{1}{n}\log\sqrt{\pi n}+O\Big(\frac{1}{n^{2}}\Big)\\
 & =\frac{1}{2}\log\frac{n}{n+2}+\frac{1}{2n}\log(\pi n)+O\Big(\frac{1}{n^{2}}\Big)\\
 & =\frac{\log n}{2n}+O\Big(\frac{1}{n}\Big),
\end{align*}
where (a) is by Stirling's approximation~\cite{jameson2015simple}. If a concrete upper bound is desired, 
\begin{align*}
 & \frac{1}{2}\log\frac{2\pi e}{n+2}-\frac{1}{n}\log\kappa_{n}\\
 & \stackrel{(b)}{\le}\frac{1}{2}\log\frac{2e}{n+2}+\frac{1}{2}\log\frac{n}{2e}+\frac{1}{n}\log\sqrt{\pi n}+\frac{1}{6n^{2}}\log e\\
 & =\frac{1}{2}\log\frac{n}{n+2}+\frac{1}{2n}\log(\pi n)+\frac{1}{6n^{2}}\log e\\
 & \le\frac{\log n}{2n}-\frac{0.05}{n}
\end{align*}
for $n\ge2$, where (b) is by Stirling's approximation~\cite{jameson2015simple}.

We now study the Shannon redundancy w.r.t. MSE of RSUQ \eqref{eq:red_mse_shannon}. From Corollary~\ref{cor:rsuq_ball}, the MSE of RSUQ is $\frac{nr^{2}}{n+2}$. Fixing the MSE to $D$, we have
\[
r=\sqrt{\frac{D(n+2)}{n}}.
\]
By Corollary~\ref{cor:rsuq_ball}, the normalized entropy is upper-bounded by
\begin{align*}
 & -n\log\sqrt{\frac{D(n+2)}{n}}-\log\kappa_{n}+\log e.
\end{align*}
Hence, the Shannon redundancy w.r.t. MSE is at most
\begin{align*}
 & \frac{1}{2}\log\frac{2\pi e}{n+2}+\frac{1}{n}\log\frac{e}{\kappa_{n}}.
\end{align*}
To obtain a simpler bound, note that the Zador redundancy w.r.t. MSE \eqref{eq:red_mse_zador} is $(\log e)/n$. Combining this with the fact that the gap between \eqref{eq:red_mse_shannon} and \eqref{eq:red_mse_zador} is upper-bounded by $(\log n)/(2n)-0.05/n$ gives the desired bound.

\subsection{Comparison of Redundancies w.r.t. MSE}  \label{subsec:NSM_detailed_plot}

\begin{figure}[htbp]
\begin{centering}
\includegraphics[scale=0.45]{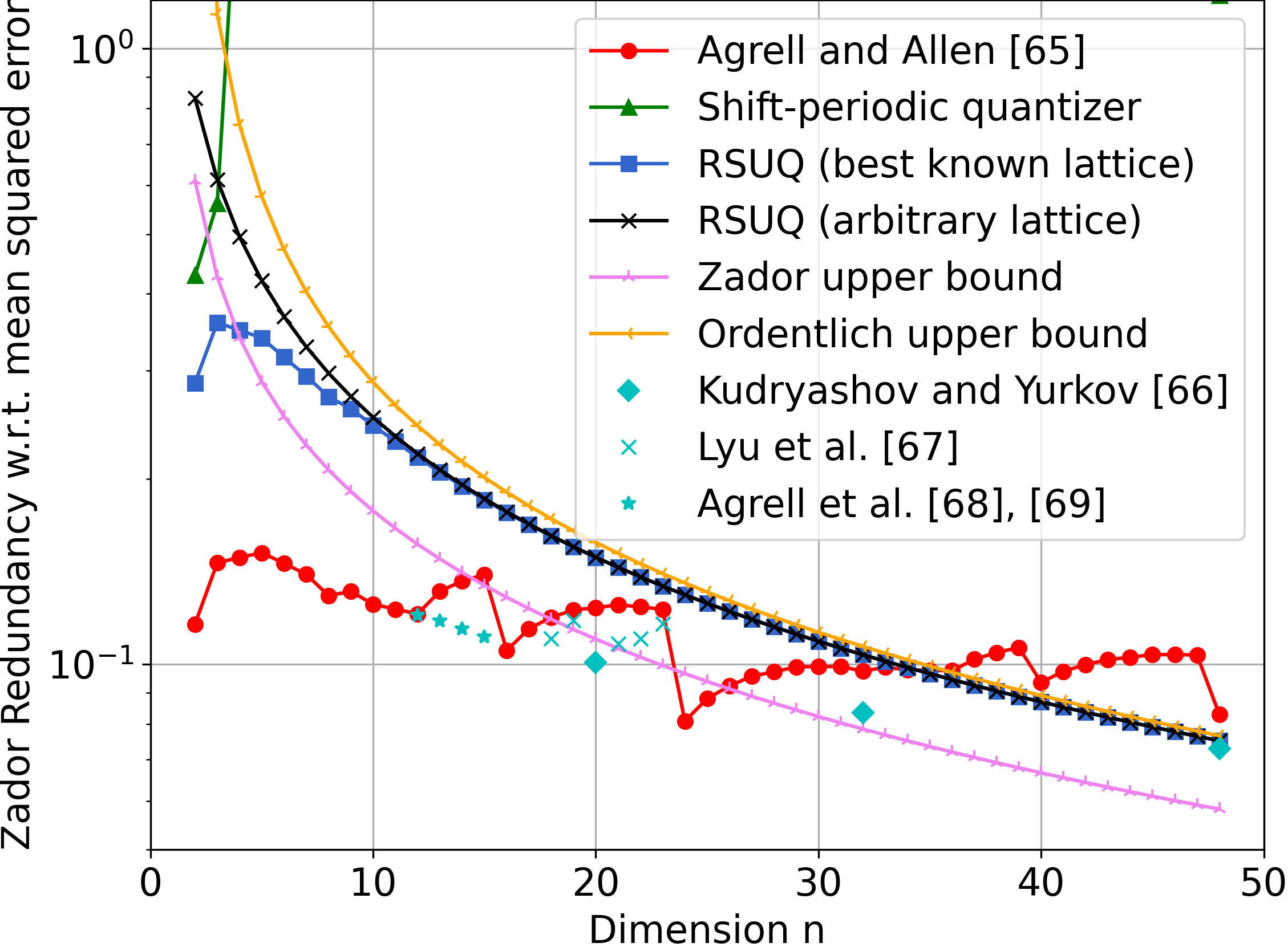}
\par
\end{centering}
\caption{\label{fig:compare_lattice1}Log-scale plot of the Zador redundancy w.r.t. MSE~\eqref{eq:lattice_mse_zador} of the lattice quantizer  with the NSM $G_n(\mathcal{V})$ from \cite{agrell2023best} (red line), from \cite{Kudryashov2010LowComplexLattice} (cyan diamond), from \cite{Lyu2022BetterLatticeQuan} (cyan cross), and from \cite{Agrell2025OptLatticeQuan} (cyan star),
the shift-periodic quantizer (green line),
the RSUQ with error distribution $\mathrm{Unif}(B^{n})$ constructed
using the best known sphere packing lattice (blue line), 
the RSUQ constructed using an arbitrary lattice (black line),  Zador's upper bound \eqref{eq:zador_ub} (violet line), and Ordentlich's upper bound \eqref{eq:or_bound} (orange line).}
\vspace{-3pt}
\end{figure}

Figure~\ref{fig:compare_lattice1} is a more detailed version of the right panel of Figure~\ref{fig:compare_lattice}, where we plot the results in \cite{agrell2023best,Kudryashov2010LowComplexLattice,Lyu2022BetterLatticeQuan,Agrell2025OptLatticeQuan} separately.
\begin{itemize}
    \item In \cite{agrell2023best}, Agrell and Allen computed the NSM of the product lattices and compiled a table of best known lattice quantizers at the time. 
    \item In \cite{Kudryashov2010LowComplexLattice}, Kudryashov and Yurkov constructed the best known lattice quantizers based on tailbiting convolutional codes for dimensions $n=20,32,48$.
    \item In  \cite{Lyu2022BetterLatticeQuan}, Lyu et al. presented the best known lattice quantizers based on the complex checkerboard lattices constructed using algebraic integers for dimensions $n = 18, 19, 22, 23$. 
    They also provided the NSM for a 21-dimensional product lattice, constructed as the product of a Kudryashov-Yurkov lattice (for $n = 20$) and a scaled integer lattice $a \mathbb{Z}$, where $a$ is chosen to satisfy certain constraints.
    \item In \cite{Agrell2024gluedLat}, Agrell et al. showed that for $n=12$, the glued lattice $D_6 \times D_6$ is a better lattice quantizer than Coxeter-Todd lattice $K_{12}$. $D_6 \times D_6$ is the current best known lattice quantizer for $n=12$ \cite{Agrell2025OptLatticeQuan}.
    \item In \cite{Agrell2025OptLatticeQuan}, Agrell et al. applied stochastic gradient descent to obtain the best known lattices for $n=13,14,15$.
\end{itemize}

\subsection{Proof of Theorem 
\ref{thm:rsuq_ent_gen} \label{subsec:pf_rsqu_ent_gen}}

Fix $S=s$, i.e., fix a sequence  $(\mathbf{V}_i)_i=(\mathbf{v}_i)_i$ of random dithers. Assume $Q_{\mathcal{A},\mathcal{P}}$ induces a partition $(\mathcal{C}_i)_i$ of $\mathbb{R}^n$ (which depends on $(\mathbf{V}_i)_i$), i.e., each $\mathcal{C}_i$ is a quantization cell $\{\mathbf{x} \in \mathbb{R}^n :\, Q_{\mathcal{A},\mathcal{P}}(\mathbf{x}, (\mathbf{v}_i)_i) = \mathbf{y}\}$ for some $\mathbf{y} \in \mathbb{R}^n$. Notice that  
\[P_\mathbf{X}(\mathcal{C}_i) = \int_{\mathcal{C}_i} f_{\mathbf{X}}(\mathbf{x})\mathrm{d}\mathbf{x},\]
and that the diameters of all $\mathcal{C}_i$'s is bounded by $\epsilon$, i.e., 
\[
\max_{i} \mathrm{diam}(\mathcal{C}_i) \leq \epsilon.
\]
We want to bound the conditional entropy
\[
H(Q_{\mathcal{A},\mathcal{P}}(\mathbf{X},s)|S=s)=\sum_i P_\mathbf{X}(\mathcal{C}_i) \log \frac{1}{P_\mathbf{X}(\mathcal{C}_i)}.
\]

For the lower bound, we have 
\begin{align*}
&\int_{\mathcal{C}_i} f_\mathbf{X}(\mathbf{x}) \log \frac{1}{f_\mathbf{X}(\mathbf{x})} \mathrm{d}\mathbf{x} \\
& =\mu(\mathcal{C}_i) \cdot \frac{1}{\mu(\mathcal{C}_i)}\int_{\mathcal{C}_i} f_\mathbf{X}(\mathbf{x}) \log \frac{1}{f_\mathbf{X}(\mathbf{x})} \mathrm{d}\mathbf{x}\\
\\ &\stackrel{(a)}{\le} \mu(\mathcal{C}_i) \left( \frac{\int_{\mathcal{C}_i} f_\mathbf{X}(\mathbf{x}) \mathrm{d}\mathbf{x}}{\mu(\mathcal{C}_i)} \right) \log \frac{1}{\Big( \frac{\int_{\mathcal{C}_i} f_\mathbf{X}(\mathbf{x}) \mathrm{d}\mathbf{x}}{\mu(\mathcal{C}_i)} \Big)} \\
&= P_\mathbf{X}(\mathcal{C}_i) \log \Big(\frac{\mu(\mathcal{C}_i)}{P_{\mathbf{X}}(\mathcal{C}_i)}\Big), 
\end{align*}
where (a) is by applying Jensen's inequality on the concave function $- x \log x$ over $(0,\infty)$.
Therefore, 
\begin{align*}
h(\mathbf{X}) &= \int_{\mathbb{R}^n} f_\mathbf{X}(\mathbf{x}) \log \frac{1}{f_\mathbf{X}(\mathbf{x})} \mathrm{d}\mathbf{x} \\
&= \sum_{i}\int_{\mathcal{C}_i} f_\mathbf{X}(\mathbf{x}) \log \frac{1}{f_\mathbf{X}(\mathbf{x})} \mathrm{d}\mathbf{x}\\
& \le \sum_i P_\mathbf{X}(\mathcal{C}_i) \log \Big(\frac{\mu(\mathcal{C}_i)}{P_{\mathbf{X}}(\mathcal{C}_i)}\Big).\\
& = H(Q_{\mathcal{A},\mathcal{P}}(\mathbf{X},s)|S=s) + \sum_i P_\mathbf{X}(\mathcal{C}_i) \log \mu(\mathcal{C}_i),
\end{align*}
or equivalently,
\begin{align}
h(\mathbf{X}) -  \sum_i P_\mathbf{X}(\mathcal{C}_i) \log \mu(\mathcal{C}_i)\leq H(Q_{\mathcal{A},\mathcal{P}}(\mathbf{X},s)|S=s). \label{eq:condent_lb_pf}
\end{align}

For the upper bound, we have
\begin{align*}
&\int_{\mathcal{C}_i} f_\mathbf{X}(\mathbf{x}) \log \frac{1}{f_\mathbf{X}(\mathbf{x})} \mathrm{d}\mathbf{x} \\
& = \int_{\mathcal{C}_i} f_\mathbf{X}(\mathbf{x}) \log \frac{1}{\Big( \frac{\int_{\mathcal{C}_i \cap \mathcal{X}} f_\mathbf{X}(\mathbf{x}') \mathrm{d}\mathbf{x}'}{\mu(\mathcal{C}_i \cap \mathcal{X})} \Big)}\mathrm{d}\mathbf{x}  \\
& \quad+ \int_{\mathcal{C}_i} f_\mathbf{X}(\mathbf{x}) \log \frac{\Big( \frac{\int_{\mathcal{C}_i \cap \mathcal{X}} f_\mathbf{X}(\mathbf{x}') \mathrm{d}\mathbf{x}'}{\mu(\mathcal{C}_i \cap \mathcal{X})} \Big)}{f_\mathbf{X}(\mathbf{x})} \mathrm{d}\mathbf{x}\\ 
& = P_{\mathbf{X}}(\mathcal{C}_i) \log \frac{\mu(\mathcal{C}_i \cap \mathcal{X})}{P_{\mathbf{X}}(\mathcal{C}_i)}  + \int_{\mathcal{C}_i} f_\mathbf{X}(\mathbf{x}) \log \frac{\int_{\mathcal{C}_i \cap \mathcal{X}} \frac{f_\mathbf{X}(\mathbf{x}')}{f_\mathbf{X}(\mathbf{x})} \mathrm{d}\mathbf{x}'}{\mu(\mathcal{C}_i \cap \mathcal{X}) } \mathrm{d}\mathbf{x}\\ 
& \stackrel{(b)}{\ge} P_{\mathbf{X}}(\mathcal{C}_i) \log \frac{\mu(\mathcal{C}_i \cap \mathcal{X})}{P_{\mathbf{X}}(\mathcal{C}_i)}  + \int_{\mathcal{C}_i} f_\mathbf{X}(\mathbf{x}) \log 2^{-\delta_{\epsilon}(\mathbf{x})} \mathrm{d}\mathbf{x}\\ 
& = P_{\mathbf{X}}(\mathcal{C}_i) \log \frac{\mu(\mathcal{C}_i)}{P_{\mathbf{X}}(\mathcal{C}_i)} - P_{\mathbf{X}}(\mathcal{C}_i) \log \frac{\mu(\mathcal{C}_i)}{\mu(\mathcal{C}_i \cap \mathcal{X})}  \\
\,&\quad- \int_{\mathcal{C}_i} f_\mathbf{X}(\mathbf{x}) \delta_{\epsilon}(\mathbf{x}) \mathrm{d}\mathbf{x},\\
& \ge P_{\mathbf{X}}(\mathcal{C}_i) \log \frac{\mu(\mathcal{C}_i)}{P_{\mathbf{X}}(\mathcal{C}_i)} - f_{\max} \cdot \mu(\mathcal{C}_i \cap \mathcal{X}) \log \frac{\mu(\mathcal{C}_i)}{\mu(\mathcal{C}_i \cap \mathcal{X})} \\
&\quad- \int_{\mathcal{C}_i} f_\mathbf{X}(\mathbf{x}) \delta_{\epsilon}(\mathbf{x}) \mathrm{d}\mathbf{x},\\
& \stackrel{(c)}{\ge} P_{\mathbf{X}}(\mathcal{C}_i) \log \frac{\mu(\mathcal{C}_i)}{P_{\mathbf{X}}(\mathcal{C}_i)} - f_{\max} \cdot \mu(\mathcal{C}_i \backslash \mathcal{X}) \log e  \\
\,&\quad- \int_{\mathcal{C}_i} f_\mathbf{X}(\mathbf{x}) \delta_{\epsilon}(\mathbf{x}) \mathrm{d}\mathbf{x},
\end{align*}
where $f_{\max} := \sup_{\mathbf{x}} f_{\mathbf{X}}(\mathbf{x})$, (b) is because the average value of $f_\mathbf{X}(\mathbf{x}')/f_\mathbf{X}(\mathbf{x})$ over $\mathcal{C}_i$ is lower bounded by the infimum $2^{-\delta_{\epsilon}(\mathbf{x})}$,
and (c) is due to $a \log ((a+b)/a) = a \log (1 + b/a) \le a (b/a) \log e = b \log e$ for $a, b>0$ by the concavity of $\log$.
Therefore,
\begin{align*}
&h(\mathbf{X})=\int_{\mathbb{R}^n} f_\mathbf{X}(\mathbf{x}) \log \frac{1}{f_\mathbf{X}(\mathbf{x})} \mathrm{d}\mathbf{x} \\
&\geq 
\sum_{i}P_{\mathbf{X}}(\mathcal{C}_i) \log \frac{\mu(\mathcal{C}_i)}{P_{\mathbf{X}}(\mathcal{C}_i)} - (\log e) f_{\max} \sum_{i} \mu(\mathcal{C}_i \backslash \mathcal{X})  \\
\,&\quad- \int_{\mathbb{R}^n} f_\mathbf{X}(\mathbf{x}) \delta_{\epsilon}(\mathbf{x}) \mathrm{d}\mathbf{x}   \\
& = \sum_{i}P_{\mathbf{X}}(\mathcal{C}_i) \log \frac{\mu(\mathcal{C}_i)}{P_{\mathbf{X}}(\mathcal{C}_i)} - (\log e) f_{\max} \mu \Big(\bigcup_{i:\, \mathcal{C}_i \cap \mathcal{X} \neq \emptyset} \mathcal{C}_i \backslash \mathcal{X} \Big) \\
\,&\quad- \int_{\mathbb{R}^n} f_\mathbf{X}(\mathbf{x}) \delta_{\epsilon}(\mathbf{x}) \mathrm{d}\mathbf{x}   \\
& \ge \sum_{i}P_{\mathbf{X}}(\mathcal{C}_i) \log \frac{\mu(\mathcal{C}_i)}{P_{\mathbf{X}}(\mathcal{C}_i)} - (\log e) f_{\max} \mu \big((\mathcal{X} + \epsilon B^n) \backslash \mathcal{X} \big) \\
\,&\quad- \int_{\mathbb{R}^n} f_\mathbf{X}(\mathbf{x}) \delta_{\epsilon}(\mathbf{x}) \mathrm{d}\mathbf{x}   \\
& =  H(Q_{\mathcal{A},\mathcal{P}}(\mathbf{X},s)|S=s) + \sum_i P_\mathbf{X}(\mathcal{C}_i) \log \mu(\mathcal{C}_i) - \overline{\delta}_{\epsilon}.
\end{align*}
Combining this with \eqref{eq:condent_lb_pf},
\begin{align*}
0 &\le H(Q_{\mathcal{A},\mathcal{P}}(\mathbf{X},s)|S=s) - \Big( h(\mathbf{X}) - \sum_i P_\mathbf{X}(\mathcal{C}_i) \log \mu(\mathcal{C}_i) \Big)  \\ 
&\le \overline{\delta}_{\epsilon}.    
\end{align*}
Taking expectation over $S = (\mathbf{V}_i)_i$,
\begin{align}
0 &\le H(Q_{\mathcal{A},\mathcal{P}}(\mathbf{X},S)|S) - \Big( h(\mathbf{X})-  \mathbb{E}\Big[\sum_i P_\mathbf{X}(\mathcal{C}_i) \log \mu(\mathcal{C}_i) \Big] \Big)  \nonumber\\
&\le \overline{\delta}_{\epsilon}. \label{eq:HhE_bd_pf}
\end{align}
To prove the theorem, it is left to show that $\mathbb{E}[\sum_i P_\mathbf{X}(\mathcal{C}_i) \log \mu(\mathcal{C}_i) ] = \overline{H}(Q)$. Let $\mathbf{X} \sim f_{\mathbf{X}}$, and define $\mathcal{C}(\mathbf{X},(\mathbf{V}_i)_i)$ to be the $\mathcal{C}_i$ that contains $\mathbf{X}$. We have
\begin{align*}
& \mathbb{E}_{(\mathbf{V}_i)_i}\Big[\sum_i P_\mathbf{X}(\mathcal{C}_i) \log \mu(\mathcal{C}_i) \Big] \\
& = \mathbb{E}_{\mathbf{X},(\mathbf{V}_i)_i}\Big[\log \mu(\mathcal{C}(\mathbf{X},(\mathbf{V}_i)_i)) \Big]\\
& = \mathbb{E}_{(\mathbf{V}_i)_i}\Big[\log \mu(\mathcal{C}(\mathbf{x},(\mathbf{V}_i)_i)) \Big],
\end{align*}
where the last equality holds for every fixed $\mathbf{x}$ since RSUQ, like the conventional subtractively dithered lattice quantization, is invariant under shifting, i.e., the RSUQ applied on $\mathbf{x}$ with the dither signals $(\mathbf{V}_i)_i$ is the same as the RSUQ applied on $\mathbf{x}'$ with the dither signals $(\mathbf{V}_i + \mathbf{x}' - \mathbf{x})_i$ (modulo the basic cell $\mathcal{P}$). Hence, 
\begin{align}
0 &\le H(Q_{\mathcal{A},\mathcal{P}}(\mathbf{X},S)|S) - \Big( h(\mathbf{X}) -  \mathbb{E}_{(\mathbf{V}_i)_i}\Big[\log \mu(\mathcal{C}(\mathbf{x},(\mathbf{V}_i)_i)) \Big] \Big) \nonumber\\ 
&\le \overline{\delta}_{\epsilon}.\label{eq:HhE_bd}
\end{align}
Note that $H(Q_{\mathcal{A},\mathcal{P}}(\mathbf{X},S)|S) - h(\mathbf{X})$ depends on the distribution of $\mathbf{X}$, whereas $\mathbb{E}_{(\mathbf{V}_i)_i}[\cdots]$ does not depend on the distribution of $\mathbf{X}$. Therefore, to find $\mathbb{E}_{(\mathbf{V}_i)_i}[\cdots]$, we can substitute another distribution of $\mathbf{X}$ and apply the inequality \eqref{eq:HhE_bd}. Since the normalized entropy $\overline{H}(Q)$ is defined over the distribution $\tilde{\mathbf{X}} \sim \mathrm{Unif}( \tau B^n)$ for $\tau > 0$, supported over $\tilde{\mathcal{X}} = \tau B^n$, we will apply the inequality \eqref{eq:HhE_bd} on this distribution. For this distribution, we have $\delta_\epsilon(\tilde{\mathbf{x}}) = 0$ for $\tilde{\mathbf{x}} \in \tilde{\mathcal{X}}$, and $\overline{\delta}_{\epsilon} \to 0$ as $\tau \to \infty$. Therefore, for $\tilde{\mathbf{X}} \sim \mathrm{Unif}( \tau B^n)$,
\begin{align*}
&\lim_{\tau \to \infty} \Big(H(Q_{\mathcal{A},\mathcal{P}}(\tilde{\mathbf{X}},S)|S)- \Big( h(\tilde{\mathbf{X}}) -  \mathbb{E}_{(\mathbf{V}_i)_i}\Big[\log \mu(\mathcal{C}(\mathbf{x},(\mathbf{V}_i)_i)) \Big] \Big)\Big) = 0.    
\end{align*}
By the definition of the normalized entropy $\overline{H}(Q)$,
\[
\overline{H}(Q) -  \mathbb{E}_{(\mathbf{V}_i)_i}\Big[\log \mu(\mathcal{C}(\mathbf{x},(\mathbf{V}_i)_i)) \Big]  = 0.
\]
Combining this with \eqref{eq:HhE_bd_pf} yields the desired result.

\subsection{Proof of Proposition \ref{prop:lim_deltabar}\label{subsec:pf_lim_deltabar}}

The condition $\lim_{\epsilon \to 0} \overline{\delta}_{\epsilon} = 0$ is satisfied if $\log f_{\mathbf{X}}(\mathbf{x})$ is Lipschitz over $\mathcal{X}$, and  $\mathcal{X}$ is $\mathbb{R}^n$ or has a measure-zero boundary. To show this, 
\begin{align*}
& \lim_{\epsilon \to 0} \mu ((\mathcal{X} + \epsilon B^n) \backslash \mathcal{X}) \\
&= \lim_{k \to \infty} \mu ((\mathcal{X} + k^{-1} B^n) \backslash \mathcal{X}) \\
&= \mu (\bigcap_{k \ge 1} (\mathcal{X} + k^{-1} B^n) \backslash \mathcal{X}) \\
&= \mu (\mathrm{cl}(\mathcal{X}) \backslash \mathcal{X}) \\
&= 0
\end{align*}
if $\mathcal{X}$ is $\mathbb{R}^n$ or has a measure-zero boundary. 

The condition $\lim_{\epsilon \to 0} \overline{\delta}_{\epsilon} = 0$ is also satisfied if $\mathbf{X}$ follows a multivariate Gaussian distribution.
To see this, assume that $\mathbf{X}\sim \mathcal{N}(\mathbf{0}, \mathbf{\Sigma})$. Then 
\begin{align*}
\delta_{\epsilon}(\mathbf{x}) &=\sup_{\mathbf{x}' \in \mathbb{R}^n:\,\Vert\mathbf{x}-\mathbf{x}'\Vert\le \epsilon}\log \frac{f_\mathbf{X}(\mathbf{x})}{f_\mathbf{X}(\mathbf{x}')} \\
&=\sup_{\mathbf{x}' \in \mathbb{R}^n:\,\Vert\mathbf{x}-\mathbf{x}'\Vert\le \epsilon} \Big(\frac{\Vert\mathbf{\Sigma}^{-1/2}\mathbf{x}\Vert^2}{2}-\frac{\Vert\mathbf{\Sigma}^{-1/2}\mathbf{x}'\Vert^2}{2}\Big) \\
&=\frac{1}{2}\sup_{\mathbf{x}' \in \mathbb{R}^n:\,\Vert\mathbf{x}-\mathbf{x}'\Vert\le \epsilon} \big(\Vert\mathbf{\Sigma}^{-1/2}\mathbf{x}\Vert+\Vert\mathbf{\Sigma}^{-1/2}\mathbf{x}'\Vert\big) \\
\,&\quad\cdot\big(\Vert\mathbf{\Sigma}^{-1/2}\mathbf{x}\Vert-\Vert\mathbf{\Sigma}^{-1/2}\mathbf{x}'\Vert\big) \\
& \le \frac{1}{2}\big(\lambda_{min}^{-1/2}(\mathbf{\Sigma})(2\Vert \mathbf{x} \Vert + \epsilon)\big)\big(\lambda_{min}^{-1/2}(\mathbf{\Sigma})\epsilon\big) \\
& = \epsilon \lambda_{min}^{-1}(\mathbf{\Sigma})(\Vert \mathbf{x} \Vert + \epsilon / 2),
\end{align*}
where the penultimate inequality follows since $\Vert \mathbf{x}'\Vert \le \Vert \mathbf{x}\Vert + \Vert \mathbf{x}' - \mathbf{x}\Vert  \le \Vert \mathbf{x}\Vert + \epsilon$ and $\Vert \mathbf{\Sigma}^{-1/2} \Vert_2 \le \lambda_{min}^{-1/2}(\mathbf{\Sigma})$.
Therefore, $\overline{\delta}_{\epsilon} = \int_{\mathcal{X}} \delta_{\epsilon}(\mathbf{x})f_{\mathbf{X}}(\mathbf{x}) \mathrm{d}\mathbf{x} \le \epsilon \lambda_{min}^{-1}(\mathbf{\Sigma})(\mathbb{E}[\Vert \mathbf{X} \Vert] + \epsilon / 2) \to 0$ as $\epsilon \rightarrow 0$.

\subsection{Proof of Theorem \ref{thm:universal}\label{subsec:pf_universal}}

The arguments in the proof are mostly similar to \cite{zamir1992universal},
which are included here for the sake of completeness. Let $\mathbf{Y}=Q_{\mathcal{P}}(\mathbf{X}-\mathbf{V}_{K})+\mathbf{V}_{K}$
be the output. Let $\mathbf{Z}=\mathbf{Y}-\mathbf{X}$. Note that
$\mathbf{Z}\sim\mathrm{Unif}(\mathcal{A})$ is independent of $\mathbf{X}$.
Fix any $P_{\hat{\mathbf{X}}|\mathbf{X}}$ satisfying $\mathbb{E}[d(\hat{\mathbf{X}}-\mathbf{X})]\le D$.
Assume $\hat{\mathbf{X}}|\mathbf{X}\sim P_{\hat{\mathbf{X}}|\mathbf{X}}$,
and $(\mathbf{X},\hat{\mathbf{X}})$ is independent of $\mathbf{Z}$.
We have 
\begin{align*}
H(\mathbf{Y}|(\mathbf{V}_{i})_{i}) & \le H(\mathbf{Y}|\mathbf{V}_{K})+H(\mathbf{V}_{K}|(\mathbf{V}_{i})_{i})\\
 & \stackrel{(a)}{\le}H(\mathbf{Y}|\mathbf{V}_{K})+H(K)\\
 & \stackrel{(b)}{=}I(\mathbf{X};\mathbf{Y}|\mathbf{V}_{K})+H(K)\\
 & \le I(\mathbf{X};\mathbf{Y},\mathbf{V}_{K})+H(K)\\
 & \stackrel{(c)}{=}I(\mathbf{X};\mathbf{Y})+H(K)\\
 & \le I(\mathbf{X};\mathbf{Y},\hat{\mathbf{X}})+H(K)\\
 & =I(\mathbf{X};\hat{\mathbf{X}})+I(\mathbf{X};\mathbf{Y}|\hat{\mathbf{X}})+H(K),
\end{align*}
where (a) is because $H(\mathbf{V}_{K}|(\mathbf{V}_{i})_{i},K)=0$,
(b) is because $H(\mathbf{Y}|\mathbf{X},\mathbf{V}_{K})=0$, (c) is
because $H(\mathbf{V}_{K}|\mathbf{Y})=0$ since $\mathbf{V}_{K}$
is the unique element in $(\mathbf{Y}+\mathbf{G}\mathbb{Z}^{n})\cap\mathcal{P}$.
We have
\begin{align*}
I(\mathbf{X};\mathbf{Y}|\hat{\mathbf{X}}) & =h(\mathbf{Y}|\hat{\mathbf{X}})-h(\mathbf{Y}|\mathbf{X},\hat{\mathbf{X}})\\
 & =h(\mathbf{Y}-\hat{\mathbf{X}}|\hat{\mathbf{X}})-h(\mathbf{Z})\\
 & \le h(\mathbf{Y}-\hat{\mathbf{X}})-h(\mathbf{Z})\\
 & =h(\mathbf{X}-\hat{\mathbf{X}}+\mathbf{Z})-h(\mathbf{Z})\\
 & =I(\mathbf{X}-\hat{\mathbf{X}};\mathbf{X}-\hat{\mathbf{X}}+\mathbf{Z})\\
 & \le C(D)
\end{align*}
since $\mathbb{E}[d(\hat{\mathbf{X}}-\mathbf{X})]\le D$. Also note that $H(K)= -\log p-\frac{1-p}{p}\log(1-p) \le -\log p + \log e$ as shown in the proof of Theorem~\ref{thm:rsuq_ent}. This completes the proof.

\subsection{Proof of Theorem \ref{thm:layered_alt}\label{subsec:pf_layered_alt}}

We show that for all jointly distributed $(\mathbf{Z},W)$ where $\mathbf{Z}\sim f$
and $P_{\mathbf{Z}|W}(\cdot|w)=\mathrm{Unif}(\mathcal{A}_{w})$ for
some $\mathcal{A}_{w}\subseteq\mathbb{R}^{n}$ with $0<\mu(\mathcal{A}_{w})<\infty$,
we have 
\[
h(\mathbf{Z}|W)=\mathbb{E}[\log\mu(\mathcal{A}_{W})]\le h_{L}(f).
\]
Let $(\mathbf{Z},T)\sim\mathrm{Unif}(\{(\mathbf{z},t):\,0\le t\le f(\mathbf{z})\})$.
We have $P_{\mathbf{Z}|T}(\cdot|t)=\mathrm{Unif}(L_{t}^{+}(f))$ and
$h(\mathbf{Z}|T)=h_{L}(f)$.

We first show the following claim: for every $\gamma\ge0$, 
\begin{equation}
\mathbb{E}\Big[\max\Big\{1-\frac{\gamma}{\mu(\mathcal{A}_{W})},\,0\Big\}\Big]\le\mathbb{E}\Big[\max\Big\{1-\frac{\gamma}{\mu(L_{T}^{+}(f))},\,0\Big\}\Big].\label{eq:layered_alt_claim}
\end{equation}
To show this claim, let $\mathcal{C}\subseteq\mathbb{R}^{n}$ be such
that $\mu(\mathcal{C})=\gamma$ and $\inf_{\mathbf{z}\in\mathcal{C}}f(\mathbf{z})\ge\sup_{\mathbf{z}\in\mathbb{R}^{n}\backslash\mathcal{C}}f(\mathbf{z})$.\footnote{We can find $\mathcal{C}$ as follows. Let $t_{0}=\inf\{t:\,\mu(L_{t}^{+}(f))\le\gamma\}$.
Although taking $\mathcal{C}=L_{t_{0}}^{+}(f)$ might not work since
$t\mapsto\mu(L_{t}^{+}(f))$ might be discontinuous, since $\mu(\bigcup_{t>t_{0}}L_{t}^{+}(f))\le\gamma$
and $\mu(L_{t_{0}}^{+}(f))\ge\gamma$, we can still take $\mathcal{C}=\bigcup_{t>t_{0}}L_{t}^{+}(f)\cup(L_{t_{0}}^{+}(f)\cap rB^{n})$,
where $r\ge0$ is chosen such that $\mu(\mathcal{C})=\gamma$.} We have
\begin{align*}
 & \mathbb{E}[\max\{1-\gamma/\mu(\mathcal{A}_{W}),\,0\}]\\
 & =\mathbb{E}\left[\frac{\max\{\mu(\mathcal{A}_{W})-\gamma,\,0\}}{\mu(\mathcal{A}_{W})}\right]\\
 & \stackrel{(a)}{\le}\mathbb{E}\left[\frac{\mu(\mathcal{A}_{W}\backslash\mathcal{C})}{\mu(\mathcal{A}_{W})}\right]\\
 & =\mathbb{E}\big[P_{\mathbf{Z}|W}(\mathcal{A}_{W}\backslash\mathcal{C}\,|\,W)\big]\\
 & =1-\mathbb{E}\big[P_{\mathbf{Z}|W}(\mathcal{C}\,|\,W)\big]\\
 & =1-P_{\mathbf{Z}}(\mathcal{C}).
\end{align*}
Equality in (a) holds if $\mu(\mathcal{A}_{W}\backslash\mathcal{C})=\max\{\mu(\mathcal{A}_{W})-\gamma,\,0\}$,
or equivalently, $\mu(\mathcal{A}_{W}\backslash\mathcal{C})=0$ or
$\mu(\mathcal{C}\backslash\mathcal{A}_{W})=0$ almost surely. This
holds when $\mathcal{A}_{W}=L_{t}^{+}(f)$ for some $t$ since $\inf_{\mathbf{z}\in\mathcal{C}}f(\mathbf{z})\ge\sup_{\mathbf{z}\in\mathbb{R}^{n}\backslash\mathcal{C}}f(\mathbf{z})$,
which is the case for $T$. Therefore, $\mathbb{E}[\max\{1-\gamma/\mu(L_{T}^{+}(f)),\,0\}]=1-\int_{\mathcal{C}}f(\mathbf{z})\mathrm{d}\mathbf{z}$,
and the claim \eqref{eq:layered_alt_claim} is proved.

For $0<\beta<\alpha$,
\begin{align*}
 & (\log e)\int_{\beta}^{\infty}\frac{1}{\gamma}\max\{1-\gamma/\alpha,\,0\}\mathrm{d}\gamma\\
 & =(\log e)\int_{\beta}^{\infty}\max\{1/\gamma-1/\alpha,\,0\}\mathrm{d}\gamma\\
 & =(\log e)\int_{\beta}^{\alpha}(1/\gamma-1/\alpha)\mathrm{d}\gamma\\
 & =(\log e)\Big(\ln\frac{\alpha}{\beta}+\frac{\beta}{\alpha}-1\Big)\\
 & =\log\frac{\alpha}{\beta}-(\log e)\max\Big\{1-\frac{\beta}{\alpha},\,0\Big\}.
\end{align*}
Hence, for any $\alpha,\beta>0$,
\begin{align*}
 & \max\{\log(\alpha/\beta),\,0\}\\
 & =(\log e)\bigg(\!\max\Big\{1-\frac{\beta}{\alpha},0\Big\}+\int_{\beta}^{\infty}\frac{1}{\gamma}\max\Big\{1-\frac{\gamma}{\alpha},0\Big\}\mathrm{d}\gamma\bigg).
\end{align*}
Applying this to \eqref{eq:layered_alt_claim}, we have, for any $\beta>0$,
\[
\mathbb{E}\Big[\max\Big\{\log\frac{\mu(\mathcal{A}_{W})}{\beta},\,0\Big\}\Big]\le\mathbb{E}\Big[\max\Big\{\log\frac{\mu(L_{T}^{+}(f))}{\beta},\,0\Big\}\Big],
\]
and hence
\begin{align*}
& \mathbb{E}\Big[\max\Big\{\log\mu(\mathcal{A}_{W}),\,\log\beta\Big\}\Big] \\
& \le\mathbb{E}\Big[\max\Big\{\log\mu(L_{T}^{+}(f)),\,\log\beta\Big\}\Big].
\end{align*}
Taking $\beta\to0$, we have $\mathbb{E}[\log\mu(\mathcal{A}_{W})]\le\mathbb{E}[\log\mu(L_{T}^{+}(f))]$.

\subsection{Proof of Theorem \ref{thm:rsuq_ent_layer}\label{subsec:pf_rsuq_ent_layer}}

Since LRSUQ can be regarded as applying RSUQ for $L_{T}^{+}(f)$ against
$\beta(T)\mathcal{P}$ with a random $T$, we can invoke Theorem \ref{thm:rsuq_ent}
to obtain
\begin{align*}
 & H(Q_{f,\mathcal{P}}(\mathbf{X},S)\,|\,S)\\
 & \leq\mathbb{E}\Big[\log\mu(\mathcal{X}+L_{T}^{+}(f)+\beta(T)\mathcal{P}-\beta(T)\mathcal{P})\\
 & \;\;\;\;\;\;\;\;-\log\mu(L_{T}^{+}(f))+\log e\Big]\\
 & \le\mathbb{E}\left[\log\mu(\mathcal{X}+3\eta\beta(T)B^{n})\right]-h_{L}(f)+\log e
\end{align*}
since $L_{T}^{+}(f)\subseteq\beta(T)\mathcal{P}\subseteq\eta\beta(T)B^{n}$.
For the normalized entropy, assume $\mathbb{E}[\log(1+\beta(T))]<\infty$
and take $\mathcal{X}=\tau B^{n}$. We have
\begin{align*}
 & \mathbb{E}\big[\log\mu(\mathcal{X}+3\eta\beta(T)B^{n})\big]-\log\mu(\tau B^{n})\\
 & =\mathbb{E}\big[\log\mu((\tau+3\eta\beta(T))B^{n})\big]-\log\mu(\tau B^{n})\\
 & =n\mathbb{E}\big[\log(1+3\eta\beta(T)/\tau)\big]\\
 & \to0
\end{align*}
as $\tau\to\infty$ by the dominated convergence theorem since $\mathbb{E}[\log(1+\beta(T))]<\infty$.

\subsection{Proof of Corollary \ref{cor:rsuq_exist}\label{subsec:pf_rsuq_exist}}

Let $\mathcal{P}$ be a basic cell such that $B^{n}\subseteq\mathcal{P}$.
Let $g(x):=\sup_{\mathbf{z}:\,\Vert\mathbf{z}\Vert\ge x}f(\mathbf{z})$.
Taking $\beta(t)=\sup\{x\ge0:\,g(x)\ge t\}$, we have $L_{t}^{+}(f)\subseteq\{\mathbf{z}:\,g(\Vert\mathbf{z}\Vert)\ge t\}\subseteq\beta(t)B^{n}\subseteq\beta(t)\mathcal{P}$.
Hence,
\begin{align*}
 & \mathbb{E}[\log(1+\beta(T))]\\
 & =\int_{0}^{\infty}\mu(L_{t}^{+}(f))\log(1+\beta(t))\mathrm{d}t\\
 & \le\int_{0}^{\infty}\mu(\beta(t)B^{n})\log(1+\beta(t))\mathrm{d}t\\
 & =\kappa_{n}\int_{0}^{\infty}(\beta(t))^{n}\log(1+\beta(t))\mathrm{d}t\\
 & \stackrel{(a)}{\le}\kappa_{n}(n + \log e)\int_{0}^{\infty}\int_{0}^{\beta(t)}x^{n-1}\log(1+x)\mathrm{d}x\mathrm{d}t\\
 & =\kappa_{n}(n + \log e)\int_{0}^{\infty}\int_{0}^{g(x)}x^{n-1}\log(1+x)\mathrm{d}t\mathrm{d}x\\
 & =\kappa_{n}(n + \log e)\int_{0}^{\infty}g(x)x^{n-1}\log(1+x)\mathrm{d}x,
\end{align*}
where (a) is because
\begin{align*}
 & \frac{\mathrm{d}x^{n}\log(1+x)}{\mathrm{d}x}\\
 & =nx^{n-1}\log(1+x)+\frac{x^{n}\log e}{1+x}\\
 & =x^{n-1}\Big(n\log(1+x)+\frac{x\log e}{1+x}\Big)\\
 & \le (n + \log e) x^{n-1}\log(1+x).
\end{align*}
The result follows from Theorem \ref{thm:rsuq_ent_layer}.

\subsection{Proof of Theorem \ref{thm:nonunif_converse}\label{subsec:pf_nonunif_converse}}

We use a simplified version of the arguments in \cite[Theorem 4]{hegazy2022randomized}.
Let $\mathbf{X}\sim\mathrm{Unif}(\mathcal{X})$. Let $\mathbf{Y}:=Q(\mathbf{X},S)$.
Write $Q^{-1}(\mathbf{y},s):=\{\mathbf{x}\in\mathbb{R}^{n}:\,Q(\mathbf{x},s)=\mathbf{y}\}$.
If we know $(\mathbf{Y},S)$, then we know $\mathbf{X}\in Q^{-1}(\mathbf{Y},S)$,
and hence the conditional distribution of $\mathbf{X}$ given $(\mathbf{Y},S)$
is $\mathrm{Unif}(Q^{-1}(\mathbf{Y},S)\cap\mathcal{X})$, and the
conditional distribution of $\mathbf{Z}:=\mathbf{Y}-\mathbf{X}$ given
$(\mathbf{Y},S)$ is $\mathrm{Unif}(\mathbf{Y}-(Q^{-1}(\mathbf{Y},S)\cap\mathcal{X}))$.
Since $\mathbf{Z}\sim f$, by Theorem \ref{thm:layered_alt}, 
\begin{align*}
h_{L}(f) & \ge h(\mathbf{Z}|\mathbf{Y},S)\\
 & =\mathbb{E}\big[\log\mu(\mathbf{Y}-(Q^{-1}(\mathbf{Y},S)\cap\mathcal{X}))\big]\\
 & =\mathbb{E}\big[\log\mu(Q^{-1}(\mathbf{Y},S)\cap\mathcal{X})\big]\\
 & =\mathbb{E}\big[\log\mu(\{\mathbf{x}\in\mathcal{X}:\,Q(\mathbf{x},S)=\mathbf{Y}\})\big]\\
 & =\log\mu(\mathcal{X})+\mathbb{E}\Big[\log\frac{\mu(\{\mathbf{x}\in\mathcal{X}:\,Q(\mathbf{x},S)=\mathbf{Y}\})}{\mu(\mathcal{X})}\Big]\\
 & =\log\mu(\mathcal{X})-H(\mathbf{Y}|S),
\end{align*}
which completes the proof.

\subsection{Proof of Proposition \ref{prop:norm_excess_additive} \label{subsec:pf_lb_normexcees}}
Consider $\mathbf{X}_{\tau}\sim\mathrm{Unif}(\tau B^n)$ independent of
$S\sim P_{S}$, and $P_{\mathbf{Y}|\mathbf{X}}$ is an additive noise channel $\mathbf{Y} = \mathbf{X} + \mathbf{Z}$.
Then,
\begin{align*}
  & H(Q(\mathbf{X}_{\tau},S)\,|\,S)-I(\mathbf{X}_{\tau};\mathbf{Y}_{\tau}) \\
  &\stackrel{(a)}{\le} H(Q(\mathbf{X}_{\tau},S)\,|\,S) - \left(h(\mathbf{X}_{\tau})-h(\mathbf{Y}_{\tau}-\mathbf{X}_{\tau})\right) \\
  &=H(Q(\mathbf{X}_{\tau},S)\,|\,S) - \left(h(\mathbf{X}_{\tau})-h(\mathbf{Z})\right)\\ 
  &= \left(H(Q(\mathbf{X}_{\tau},S)\,|\,S)-\log \mu(\tau B^n)\right)+h(\mathbf{Z}),
\end{align*}
where (a) is because  
\begin{align*}
    I(\mathbf{X};\mathbf{Y})&=h(\mathbf{X})-h(\mathbf{X}|\mathbf{Y})\\
    &=h(\mathbf{X})-h(\mathbf{X}-\mathbf{Y}|\mathbf{Y})\\
    &=h(\mathbf{X})-h(\mathbf{Z}|\mathbf{Y}) \\
    &\ge h(\mathbf{X})-h(\mathbf{Z}).
\end{align*}
By taking the limit superior as $\tau \to \infty$ and dividing by $n$, we obtain
\begin{align*}
    &\frac{1}{n}\underset{\tau\to\infty}{\lim\sup}\big(H(Q(\mathbf{X}_{\tau},S)\,|\,S)-I(\mathbf{X}_{\tau};\mathbf{Y}_{\tau})\big) \\
    &\le \frac{1}{n}\left(\underset{\tau\to\infty}{\lim\sup}\big(H(Q(\mathbf{X}_{\tau},S)\,|\,S)-\log\mu(\tau B^n)\big) + h(\mathbf{Z})\right) \\
    &\stackrel{(b)}{=} \frac{1}{n}\big(\overline{H}(Q) + h(\mathbf{Z}) \big),
\end{align*}
where (b) is by Definition~\ref{def:norment}.


For the other direction, assume $v := \mathbb{E}[\Vert \mathbf{Z}\Vert^2]$ is finite. Fix $r>0$ and let $\epsilon := \mathbb{P}(\Vert \mathbf{Z}\Vert > r)$.
We have
\begin{align*}
h(\mathbf{Y}_\tau) & \le h(\mathbf{Y}_\tau, \mathbf{1}\{ \Vert \mathbf{Z}\Vert > r \}) \\
& = H_b(\epsilon) + (1-\epsilon) h(\mathbf{Y}_\tau \,|\, \Vert \mathbf{Z}\Vert \le r ) \\ 
\,&\quad+ \epsilon h(\mathbf{Y}_\tau \,|\, \Vert \mathbf{Z}\Vert > r ) \\
&\le H_b(\epsilon) + (1-\epsilon) \log \mu ((\tau + r)B^n) \\ 
\,&\quad+ \frac{n\epsilon}{2} \log \Big(2 \pi e n^{-1} \mathbb{E}[\Vert\mathbf{Y}_\tau\Vert^2 \,|\, \Vert \mathbf{Z}\Vert > r ]\Big) \\
&\le H_b(\epsilon) + (1-\epsilon) \log \mu ((\tau + r)B^n) \\ 
\,&\quad+ \frac{n\epsilon}{2} \log (2 \pi e n^{-1} v/\epsilon) \\
&= H_b(\epsilon) + (1-\epsilon)\left( \log \mu (\tau B^n) + n\log\frac{\tau + r}{ \tau} \right) \\
\,&\quad+ \frac{n\epsilon}{2} \log (2 \pi e n^{-1} v/\epsilon).
\end{align*}

Take $r = \sqrt{\tau}$. Then we have $\epsilon \to 0$ as $\tau \to \infty$, and 
\begin{align*}
& n\Phi(Q) \\
& =  \underset{\tau\to\infty}{\lim\sup}\left(\big(H(Q(\mathbf{X}_{\tau},S)\,|\,S)-I(\mathbf{X}_{\tau};\mathbf{Y}_{\tau})\big)\right) \\
& =  \underset{\tau\to\infty}{\lim\sup}\left(\big(H(Q(\mathbf{X}_{\tau},S)\,|\,S)-h(\mathbf{Y}_{\tau}) + h(\mathbf{Z})\big)\right) \\
&\ge \underset{\tau\to\infty}{\lim\sup}\bigg(H(Q(\mathbf{X}_{\tau},S)\,|\,S) - H_b(\epsilon) \\ 
\,&\quad- (1-\epsilon)\left( \log \mu (\tau B^n) + n\log\frac{\tau + \sqrt{\tau}}{ \tau} \right) \\ \,&\quad- \frac{n\epsilon}{2} \log (2 \pi e n^{-1} v/\epsilon)+h(\mathbf{Z}) \bigg)\\
&\ge \underset{\tau\to\infty}{\lim\sup}\left(H(Q(\mathbf{X}_{\tau},S)\,|\,S) - (1-\epsilon) \log \mu (\tau B^n) \right) + h(\mathbf{Z})\\ 
\,&\quad-\underset{\tau\to\infty}{\lim\sup}\bigg(H_b(\epsilon)+(1-\epsilon)n\log\frac{\tau + \sqrt{\tau}}{ \tau}\\ 
\,&\quad+\frac{n\epsilon}{2} \log (2 \pi e n^{-1} v/\epsilon)\bigg)\\
&= \overline{H}(Q) + h(\mathbf{Z}).
\end{align*}
This completes the proof.

\subsection{Layered Entropy of Multivariate Gaussian Distribution \label{subsec:layeredEnt_Gaussian}}

When the error vector $\mathbf{Z}$ follows a zero-mean  multivariate-Gaussian distribution with an identity covariance matrix $\mathbf{I}_{n \times n}$, i.e.,  $\mathbf{Z} \sim \mathcal{N}(\mathbf{0},\mathbf{I}_{n \times n})$, $\mathbf{Z}$ can be expressed as a mixture of uniform distributions over $n$-dimensional balls, with the latent variable following a chi-squared distribution with $n+2$ degrees of freedom, i.e., $\mathbf{Z} \, |\, \{V=v\} \sim \mathrm{Unif}(\sqrt{v} B^{n})$ and $V \sim \chi^{2}_{n+2}$. See Appendix~\ref{subsec:pf_Nor_mix_chi2} for the proof.\footnote{This fact was pointed out to the authors by Iosif Pinelis in \cite{Iosif2024}.}
Its layered entropy can be expressed as follows:
\begin{align*}
h_{L}(f_{\mathbf{Z}}) &= \int_{0}^{\infty}\mu(L_{t}^{+}(f_{\mathbf{Z}}))\log\mu(L_{t}^{+}(f_{\mathbf{Z}})) \mathrm{d}t \\
& = \mathbb{E}\big[ \log\mu(\sqrt{V}B^n) \big] \\
& = \int_0^{\infty} \frac{1}{2^{n/2+1} \Gamma(n/2+1)} v^{n/2} e^{-v/2} \\ 
\,&\quad\cdot\log \Big(v^{n/2} \frac{\pi^{n/2}}{\Gamma(n/2+1)}\Big) \mathrm{d}v \\
& = \int_0^{\infty} \Big(\frac{v}{2}\Big)^{n/2} e^{-v/2} \\
\,&\quad\cdot \frac{(n/2) \log(\pi v) - \log \Gamma(n/2+1)}{2 \Gamma(n/2+1)}  \mathrm{d}v.
\end{align*}

\subsection{Expressing a Multivariate Gaussian Distribution as a Mixture of Uniform Distributions over $n$-Balls\label{subsec:pf_Nor_mix_chi2} }

By the construction of LRSUQ, we can generate $\mathbf{Z} \sim N(\mathbf{0}, \mathbf{I})$ by first generating $T$ with a probability density function $f_{T}(t) := \mu(L_{t}^{+}(f))$, where $f=f_{\mathbf{Z}}$ is the probability density function of $\mathbf{Z}$, and then generate $\mathbf{Z} \in L_{t}^{+}(f)$ uniformly. Let $v \ge 0$ be such that
\[
t = (2\pi)^{-n/2} e^{-v/2}.
\]
We have $L_{t}^{+}(f) = \sqrt{v} B^n$. Hence,
\begin{align*}
f_{T}(t) &= \mu(L_{t}^{+}(f))\\
 &= v^{n/2} \mu(B^n)\\
& = v^{n/2} \frac{\pi^{n/2}}{\Gamma(n/2+1)}.
\end{align*}
We have
\begin{align*}
f_{V}(v) &= f_{T}((2\pi)^{-n/2} e^{-v/2}) \cdot \Big|\frac{\mathrm{d} }{\mathrm{d} v} (2\pi)^{-n/2} e^{-v/2} \Big| \\
& = v^{n/2}\frac{\pi^{n/2}}{\Gamma(n/2+1)} \cdot \frac{1}{2}(2\pi)^{-n/2} e^{-v/2}\\
& = \frac{v^{n/2} e^{-v/2}}{2^{(n+2)/2} \Gamma((n+2)/2)},
\end{align*}
which is the probability density function of the chi-square distribution with $n+2$ degrees of freedom.

\iflongver

\bibliographystyle{IEEEtran}
\bibliography{ref}

\fi

\fi

\end{document}